\documentclass[aps,pra,reprint,floatfix]{revtex4-2}
\usepackage{mymacros}

\usepackage[T1]{fontenc}
\usepackage[utf8]{inputenc}
\usepackage{graphicx}
\graphicspath{{figures/}}

\usepackage[dvipsnames]{xcolor}
\usepackage{subcaption}

\usepackage{bm}
\usepackage{caption}
\usepackage{amsmath, amsthm, amssymb, amsfonts}
\usepackage{bbm}
\usepackage{slashed}
\usepackage{physics}
\usepackage{epsfig}
\usepackage{xspace}
\usepackage{graphics}
\usepackage{slashed}
\usepackage{bbold}
\usepackage{bbm}
\usepackage{empheq}
\usepackage{mathrsfs}

\usepackage{hyperref}
\hypersetup{
    colorlinks=true,
    allcolors = blue
    }

\usepackage[capitalise]{cleveref}

\newcommand{\bq}{\ensuremath{{q}}}

\newcommand{\Lphy}{\ensuremath{L_{\rm phy}}}
\newcommand{\mphy}{\ensuremath{m_{\rm phy}}}

\newcommand{\Vlat}{\ensuremath{V}}

\newcommand{\Eu}{\ensuremath{E^{(\uparrow)}}}
\newcommand{\Ed}{\ensuremath{E^{(\downarrow)}}}

\newcommand{\Es}{\ensuremath{E^{(\uparrow\!\downarrow)}}}
\newcommand{\Ep}{\ensuremath{E^{(\uparrow\!\uparrow)}}}
\newcommand{\Ethrees}{\ensuremath{E^{(\uparrow\!\uparrow\!\downarrow)}}}
\newcommand{\Ethreep}{\ensuremath{E^{(3\uparrow)}}}
\newcommand{\Efourp}{\ensuremath{E^{(4\uparrow)}}}
\newcommand{\Efour}{\ensuremath{E^{(\uparrow\uparrow\uparrow\downarrow)}}}

\newcommand{\Efivep}{\ensuremath{E^{(5\uparrow)}}}

\newcommand{\orcid}[1]{\href{https://orcid.org/#1}{\includegraphics[width=8pt]{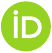}}}

\begin{document}

\title{Lattice Regularization of Non-relativistic Interacting Fermions in One Dimension }

\author{Zihan Li\orcid{0009-0007-8230-3097}}
\email{lizihan8@msu.edu}
\author{Son T. Nguyen\orcid{0000-0002-6104-7035}}
\email{snguyen@wlu.edu}
\affiliation{Department of Physics and Engineering, Washington and Lee University, Lexington 24450, USA}


\date{August 28, 2026}






\begin{abstract}

Few-body systems in one spatial dimension provide an ideal setting for testing lattice methods, since exact results are available against which lattice systematics can be assessed. We study two-component non-relativistic fermions with attractive contact interactions on a lattice, where the even ($s$-wave) interaction acts between opposite spins and the odd ($p$-wave) interaction between like spins, using exact diagonalization. The couplings are fixed entirely from two-body data. We demonstrate that the two-body couplings alone determine the few-body physics. We compute the three- and four-body systems' ground states when $s$- and 
$p$-wave interactions coexist and find that both of them, expressed relative to the two-body threshold, depend only on a single dimensionless parameter.
\end{abstract}

\maketitle

\section{Introduction}
Ultracold Fermi gases provide an excellent platform to study many-body physics due to their high degree of tunability through Feshbach resonances \cite{RevModPhys.82.1225}. They have made strongly interacting quantum systems directly accessible in experiment \cite{O_Hara_2002, PhysRevLett.94.210401, Liao_2010, Pagano_2014, Wenz_2013, PhysRevLett.102.230402,jackson2023emergent, PhysRevLett.125.263402, Venu_2023}. In particular, tight optical confinement freezes out the transverse motion of a three-dimensional gas, producing arrays of effectively one-dimensional tubes in which higher-dimensional physics can be studied in a controlled setting. The broader landscape of one-dimensional quantum gases, including bosonic realizations and multicomponent systems, is reviewed in, e.g., Refs. \cite{Bloch_2008,Cazalilla_2011,Guan_2013}.

In the low-energy limit, the physics of interacting fermions becomes universal, as the detailed microscopic structure of the interaction is integrated out \cite{Tan_2008, Barth_2011, He_2016}. For two-component fermions in one dimension (1D), this universality leads to the Gaudin-Yang model \cite{Yang:1967bm, GAUDIN196755}, which provides the continuum description of fermions interacting via a delta potential. The ground-state energy, excitation spectrum, and thermodynamic properties of this model in the continuum can be obtained exactly using the Bethe ansatz. These exact solutions have played a central role in establishing our understanding of correlation effects, pairing phenomena, and crossover from the Bardeen-Cooper-Schrieffer (BCS) state of weakly-correlated pairs of fermions to the Bose-Einstein condensation (BEC) of diatomic molecules in one-dimensional fermionic systems (see e.g., Ref.~\cite{Guan_2013} for a comprehensive review). While the Bethe ansatz provides an exact solution for the model, it relies on the assumption of non-diffractive, factorized scattering. In the presence of mass imbalance, this assumption is generically violated, leading to diffractive multi-particle scattering and the breakdown of integrability \cite{baxter2007exactly,giamarchi2004quantum, Pecak:2017mqz}. 

For identical fermions (spinless), the situation differs at the outset. Antisymmetry causes the wavefunction to vanish at coincidence, so the leading interaction is instead the odd-wave, or $p$-wave, channel, represented by a $\delta'$-type pseudopotential. One of the most striking properties of this system is Bose-Fermi correspondence. In Ref.~\cite{GAUDIN196755}, Girardeau discovered that a hardcore boson system maps directly to a system of free spinless fermions, $\Psi_F(x_1,x_2,\cdots, x_N) =\prod_{i<j}
{\rm sgn}(x_{i}-x_j)\Psi_B( x_1,x_2,\cdots, x_N)$. This mapping ensures the antisymmetry of fermions under exchange of spatial coordinates while maintaining $|\Psi_B|^2=|\Psi_F|^2$ everywhere. Therefore, the two systems then share identical spectra and density correlations. The correspondence extends beyond the hardcore constraint.  The Cheon-Shigehara model of spinless fermions interacting via an $p$-wave contact interaction \cite{Cheon_1998, Cheon:1998iy} is dual to the Lieb–Liniger model of bosons with a $s$-wave contact interaction \cite{1963PhRv..130.1605L}. The coupling strength of the $s$-wave bosonic interaction is inversely proportional to the coupling strength of the fermionic $p$-wave interaction. For the attractive interaction, the Lieb–Liniger model's $N$-body binding energy is solved by McGuire \cite{McGuire_1964} so that the ratio of the $N$-body to the two-body binding energy is a pure number independent of the interaction strength.

The duality nevertheless conceals the fact that two theories with identical spectra do not require the same operator content. Reference~\cite{PhysRevA.103.043307} showed that, although the field theory of the dual one-dimensional Bose gas only involves the $s$-wave two-body contact, a marginally relevant three-body contact term is required in the corresponding theory of spinless fermions with an $p$-wave interaction to cancel the ultraviolet divergence. Interestingly, its value introduce now new scale \cite{PhysRevA.103.043307}, which is a consequence of the fact that both Lieb–Liniger and Cheon–Shigehara models correspond to an infinite three-body scattering length, at which the three-body term remains scale invariant \cite{Tanaka_2022}. On the other hand, this term can still be tuned to a finite three-body scattering length which will break the scale invariance and introduce an explicit three-body scale (see, e.g., Refs.~\cite{Tanaka_2022,PhysRevA.97.061605, PhysRevA.97.061603}). These physical systems are different from the one considered here. 

The status of the three-body operator is not merely a formal question. For example, references~\cite{Guo:2022cyw, Guo:2022xhc,Guo:2023cfq} studied in-medium two- and three-body clusters in one-dimensional fermions, first with $p$-wave and then with coexistent $s$- and $p$-wave interactions. In these variational calculations, a finite momentum cutoff is retained by choice, and the resulting phase boundaries shift when the cutoff is changed. The three-body coupling can restore cutoff-independence.

A lattice formulation offers an alternative regularization and system framework for nonperturbative calculations of few-body observables.  The lattice spacing acts as the ultraviolet cutoff.  The couplings then depend on it. Discretization constrains the operator content. For example, identical fermions cannot occupy the same site, so the on-site three-body contact founded the bosonic theory is absent. The minimal three-body operator necessarily acts across three neighboring sites. Still on the lattice formulation of one-dimensional zero-range bosons, Ref.~\cite{Klein:2018iqa} discussed that the three-body term is unambiguously a Symanzik-type improvement operator \cite{Symanzik:1983dc,Symanzik:1983gh}, since the bosonic continuum theory is finite without it. 

In this work, we construct a lattice regularization of the one-dimensional $s$- and $p$-wave fermions in vacuum. The lattice couplings are fixed from two-body data using the analytical relation between finite- and infinite-length two-body energies.  Once the Hamiltonian and the coupling are specified, we can compute the three-, four-, and five-body ground state using exact diagonalization method. With only the even-channel interaction between distinguishable species, no genuine few-body bound state exists beyond the two-body dimer. Meanwhile, with only the odd-channel interaction between identical fermions, the universal continuum few-body physics obtained by McGuire is reproduced from the two-body coupling alone. Including a three-body operator changes the leading lattice error from $\mathcal{O}(a)$ to $\mathcal{O}(a^2)$ without altering the continuum limit.  We then extend the calculation to the two-component system, where both interactions are present, and show that the $2\!\uparrow +1\!\downarrow$ trimer and $3\!\uparrow +1\!\downarrow$ tetramer only depends on the ratio of the two two-body binding energies.

This paper is organized as follows. In \cref{sec:model} we introduce the lattice model and its interaction structure. \cref{sec:fixing_2B_couplings} describes how the two-body couplings are fixed on the lattice; there we also derive the exact relation between the finite- and infinite-length two-body energies, and validate the lattice against it. \cref{sec:Infinite Volume Extrapolation} presents our main results for the few-body spectra, first with the $s$-wave interaction alone and then with the 
$p$-wave interaction alone, benchmarked against the exactly known values. \cref{sec:universality} extends the calculation to the three-body system with both 
$s$- and $p$-wave interactions present. \cref{sec:conclusion} is a brief conclusion of our work, followed by two appendices.

\section{Lattice Model \label{sec:model}}
Our lattice model is constructed in the Fock space
formulation, using fermionic annihilation and creation operators $\psi_{\sigma,x}$ and $\psi^\dagger_{\sigma,x}$ where $x$ is the lattice site on a periodic one-dimensional spatial lattice with $L$ sites, and $\sigma=\uparrow,\downarrow$ are the two different fermion species. For simplicity, we assume they have equal masses. The lattice Hamiltonian can be arranged in two parts: a free Hamiltonian representing the kinetic part, $H_0$, and interaction terms,
\begin{equation} \label{eq:fullHlambda}
    \begin{aligned}
        H=
        H_0 +H_s+H_p,
    \end{aligned}
\end{equation}
where
    \begin{align}
       & H_0
        =\epsilon(a)\sum_{\sigma, x}\ (2n_{\sigma, x}-\psi_{\sigma, x}^\dagger\psi_{\sigma, x+1}-\psi_{\sigma, x+1}^\dagger\psi_{\sigma, x}),\\
       & 
      H_s= \epsilon(a) U(a)\sum_{x}\ n_{\uparrow,x} n_{\downarrow,x},        \label{eq:s_hamiltonian}\\
   & H_{p} =\epsilon(a)\Vlat(a)\sum_{\sigma, x} n_{\sigma, x}n_{\sigma, x+1},       \label{eq:c_hamiltonian}
    \end{align}
where $a$ is the lattice spacing, $n_{\sigma,x}=\psi_{\sigma,x}^\dagger\psi_{\sigma,x}$. The term $H_s$ acts between two distinguishable particles $(\uparrow\!\downarrow)$ on the same lattice site while the term $H_p$ acts on identical pairs ($\uparrow\!\uparrow$ or $\downarrow\!\downarrow$) on adjacent sites. In these above equations $\epsilon$ sets the overall energy scale so that $U$ and $V$ can be reported as dimensionless quantities. We assume that this model on a 1D
lattice with $L$ sites describe a continuum physical system in a periodic 1D box of physical
length $\Lphy$. The lattice spacing $a$ is
given by $a= \Lphy/L $.

First, $\epsilon(a)$ is determined by setting the ground state energy of a free particle on a lattice, $\Eu_{1,L}=\Ed_{1,L} = E_1$, where $E_1$ is the lowest non-zero energy of a single free particle in a periodic one-dimensional box of length $\Lphy$. The ground state energy of this continuous theory has the form
\begin{equation}
    E_1 = \frac{2\pi^2}{m_{\rm phy}L^2_{\rm phy}}.
\end{equation}
In what follows, we append the subscript ``phy'' to physical parameters to distinguish them from lattice parameters. It is straightforward to obtain the lattice model's energy spectrum by diagonalizing $H_0$ in momentum space, yielding
\begin{equation}
    \Eu_{n,{L}} =2\epsilon(a)\left[1-\cos\left(\frac{2\pi a}{\Lphy}n\right)\right] 
    \label{eq:single_lambda_lat}
\end{equation}
Finally, setting $\Eu_{1,L} =\Ed_{1,L}= E_1$ gives us the relationship between $\epsilon(a)$ and known variables
\begin{equation}\label{eq:B28}
    \epsilon(a)=\frac{\pi^2}{\mphy\Lphy^2}\left[1-\cos\left({\frac{2\pi a}{\Lphy}}\right)\right]^{-1}.
\end{equation}

\renewcommand{\arraystretch}{1.2}
\setlength{\tabcolsep}{10pt}
\setlength{\arrayrulewidth}{0.2mm}

\begin{table}
\centering
\begin{tabular}{cp{4.7cm}}
\hline
Notation & Description \\
\hline
$E^{(n\uparrow m\downarrow)}_L,E^{(n\uparrow)}_L$         & Energy in the continuum limit ($a\to0$) at finite box length $\Lphy$ \\
$E^{(n\uparrow m\downarrow)}_\infty,~E^{(n\uparrow)}_\infty$ & Energy extrapolated to infinite length ($\Lphy\to\infty$) at finite lattice spacing $a$ \\
$E^{(n\uparrow m\downarrow)},E^{(n\uparrow)}$           & Energy in both the continuum and infinite-length limits \\
\hline
\end{tabular}
\caption{Notation convention used throughout this paper for  energies and other observables at the three
combinations of finite/removed regulators. $n$ and $m$ stands for the number of particles.}
\label{tab:notation}
\end{table}

In the following sections, we adopt representative numerical values listed in \cref{tab:input} for the particle mass, box length, and finite-length two-body binding energy to tune the couplings through renormalization as discussed in \cref{sec:fixing_2B_couplings}. 
\begin{table}[ht!]
\centering
\begin{tabular}{ccc}
\hline
$\mphy$ & $\Lphy$ & $\Es_{L},\Ep_{L}$  \\
\hline
940 [MeV] & 3.4 [fm] & $-10$ [MeV] \\
\hline
\end{tabular}
\caption{Input parameters used in the test calculation.}
\label{tab:input}
\end{table}
These have no physical significance in themselves. These unphysical regularization scales are automatically removed in the continuum and infinite length limits. The two-body binding energies in the $s$- and $p$-wave channels will set the physical scales.  Moreover, in \cref{sec:Infinite Volume Extrapolation,sec:universality} the observables we report are the dimensionless  ratios of few-body to two-body binding energies and hence independent even of the remaining overall scale. We confirmed this explicitly by repeating the analysis with randomized values of the mass, length, and two-body energies, which leaves the extracted ratios unchanged.

\section{Two-body couplings\label{sec:fixing_2B_couplings}}
By construction an $\uparrow\!\downarrow$ pair interacts through the $s$-channel, an $\uparrow\!\uparrow$ pair through the $p$-channel. since parity is exact in 1D, the two never mix, and 
$U(a)$ and 
$V(a)$ can be tuned independently against the two-body spectrum in each channel.

\subsection{Two-body \texorpdfstring{$s$}{s}-wave channel \label{sec:s_renorm}}
In the naive continuum limit, this term becomes a delta-function interaction, $U_s\delta(x)$. Starting from the lattice Hamiltonian, and taking the lattice spacing $ a \to 0 $ with the spatial coordinate $ xa \to  x $, the interaction term becomes
\begin{align}
   \frac{U_s }{a}\sum_x n_{\uparrow,x} n_{\downarrow,x} \rightarrow U_s \int dx\, n_{\uparrow}(x)n_{\downarrow}(x),
    \label{eq:Hubbard_term}
\end{align}
where $n_\sigma(x)=\psi_\sigma^\dagger(x)\psi_\sigma(x)$. The coupling $U_s$ is a constant coefficient because, in the one-dimensional continuum, the contact interaction is ultraviolet (UV)- finite and requires no regularization. On the lattice, the lattice spacing acts as a UV regulator. At fixed $a$, the bare coupling is independent of the finite size of the space. As a first check that the renormalization is correctly implemented, we must verify that $a\epsilon(a) U(a)$ approaches $U_s$ in the continuum limit. 

We can find $U(a)$ analytically using the method outlined in Ref.~\cite{PhysRevC.110.024002}. We introduce a relative momentum $q$ and relative position $\delta$ to reduce a two-body system into a one-body system. A complete set of two particles can be defined as,
\begin{equation}
    \ket{q, \delta} = \frac{1}{\sqrt{L}}\sum_{x}\psi_{\uparrow,x}^\dagger\psi_{\downarrow,x+\delta}^\dagger e^{i(2\pi/L)qx}\ket{0}.
\label{eq:1B_basis}
\end{equation}
The action of the Hamiltonian in Eq.~\eqref{eq:fullHlambda} on this state is given by
\begin{align}
   H_0\ket{q,\delta}&= \epsilon(a) \big[4\ket{q,\delta}-z_q\ket{q,\delta+1}-z_q^*\ket{q,\delta-1}\big],\nonumber\\
H_s\ket{q,\delta}&=\epsilon(a) U(a)\delta_{\delta,0}\ket{q,\delta},
\end{align}
where $z_q=1+e^{i(2\pi/L)q}$. In this basis, the Hamiltonian is clearly block diagonal in $\ket{q}$. Since the ground state will appear in the sector with $q=0$, we can focus on that sector and denote $\ket{\delta}\equiv\ket{0,\delta}$ for short.

\begin{figure*}[t]
    \centering
     \includegraphics[width=0.47\linewidth]{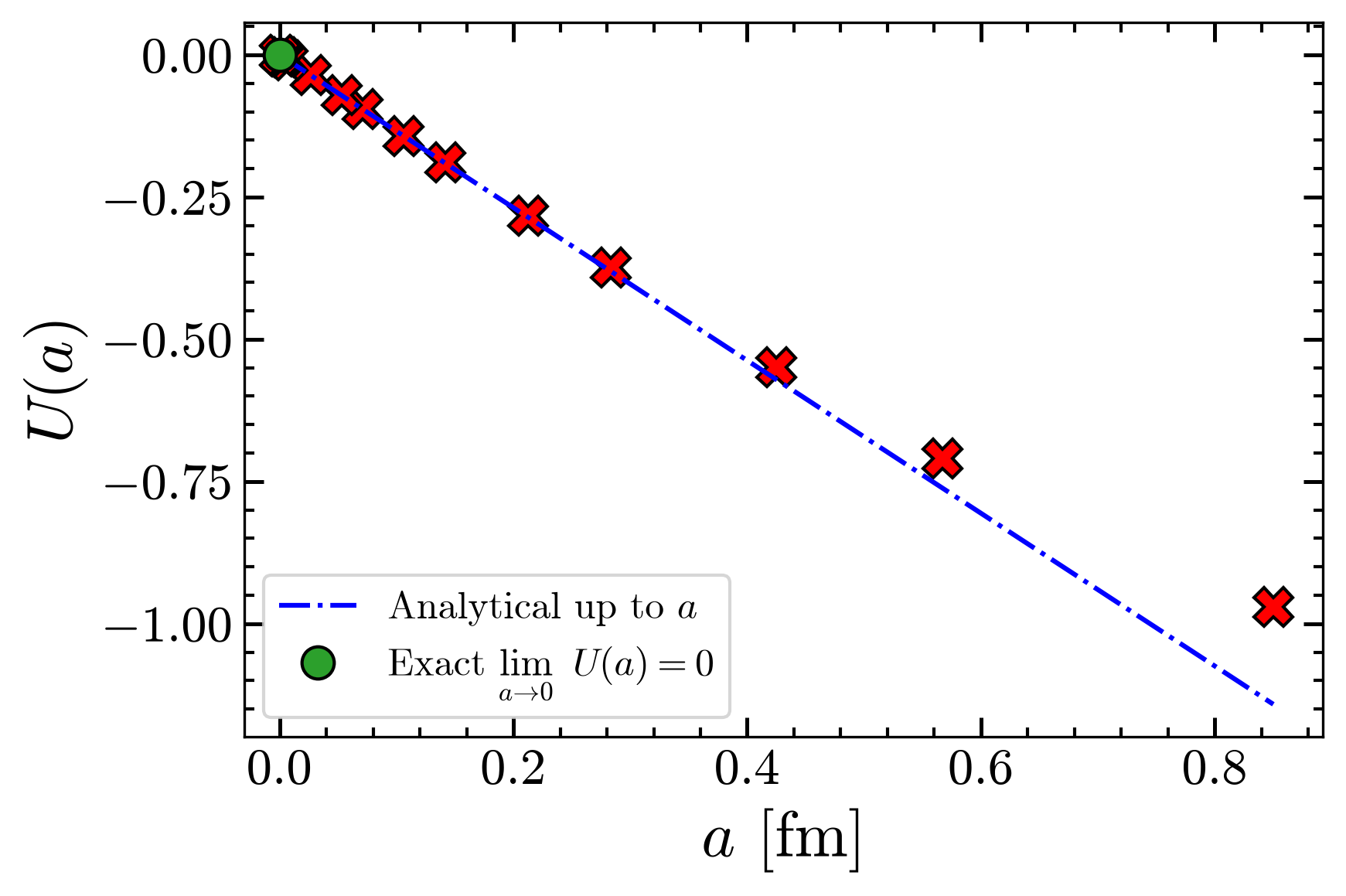} \hspace{3mm}
     \includegraphics[width=0.45\linewidth]{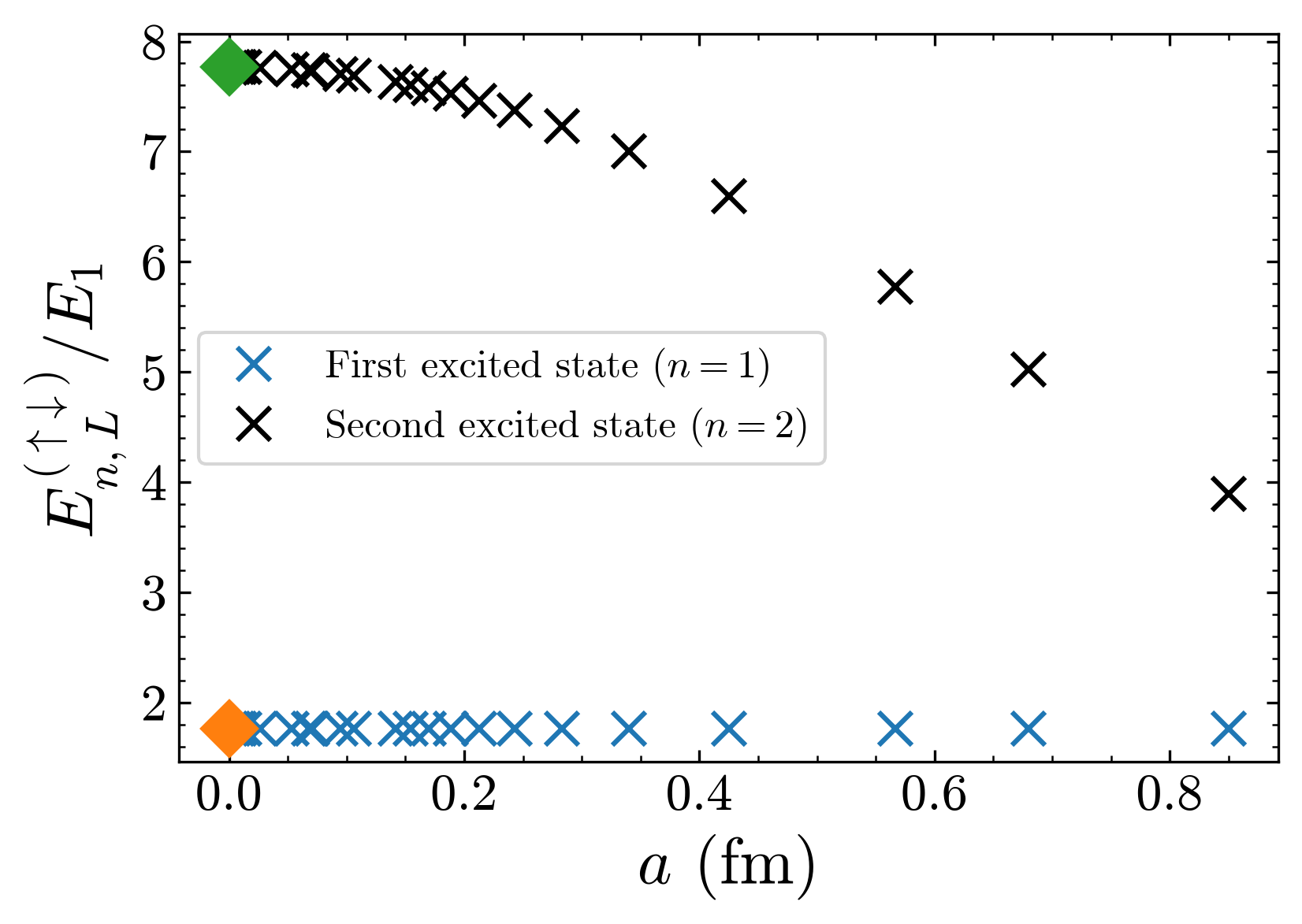}
    \caption{Left: the plot of $U(a)$ as a function of the lattice spacing using \cref{eq:B32}. Right: The plot of the ratio $\Es_{n,L}/\Eu_{1}$ for the two excited states $n= 1, 2$ as a function of the lattice spacing. The diamond symbols represent their values at the continuum, using the finite length scheme discussed in Section VI of Ref.~\cite{Korber:2019cuq} with the same testing parameters from \cref{tab:input}. The lattice sizes used in the calculations is up to $L=8192$. The dashed lines are given by \cref{eq:Ulatt_continuum}.}
    \label{fig:lambda_renorm}
\end{figure*}

As in the continuous theory, we are interested in the pole of the system. The information about potential poles is stored in the Green function matrix $G_s(E, a)=(E-H_0-H_s)^{-1}$. To simplify the full Green function, we expand it with a free Hamiltonian's Green function matrix, $G_0=(E-H_0)^{-1}$, via
\begin{align}
G_s&(E,a)=G_0(E,a)\nonumber\\
&+2\epsilon(a)\ \delta_{\delta,0}\ \mathcal{G}_s(E,a) \ G_0(E,a)|\delta\rangle\langle\delta|G_0(E,a),
\end{align}
where $\mathcal{G}_s=[2/U(a)-I_s(E,a)]^{-1}$. The dimensionless lattice interaction parameter $ U (a)$ corresponds to $ 2/I_s $,
where  $I_s$  is given by
\begin{equation} \label{eq:B32}
    I_s(E,a) = \frac{1}{L} \sum_{n=0}^{L-1} \frac{1}{E/2\epsilon(a) - 2[1 - \cos{(2\pi n / L)}]}.
\end{equation}
Since we are interested in the bound state, $E=\Es_L$ is negative. Using the Poisson summation formula (see Appendix \ref{app:Ilats} for more detail), it is possible to show that for $a\ll 1$ while $La=\Lphy$ is fixed,
\begin{align}
    I_s(a) = \frac{-1}{2a\gamma_L^{(s)}}\coth\left(\frac{\gamma_L^{(s)}\Lphy}{2}\right),
\end{align}
where $\gamma^{(s)}_L=\sqrt{\mphy|\Es_{L}|}$ is the binding momentum defined from the finite-length energy. Thus,
\begin{align}
U(a)=&-4a\gamma_L^{(s)}\tanh\left(\frac{\gamma_L^{(s)}\Lphy}{2}\right),
  \label{eq:Ulatt_continuum}
\end{align}
As $a\to 0$  we find that $ \lim_{a\to 0}U(a) = 0$. We can verify this value via exact diagonalization. We increase the number of lattice sites $L$ up to 8192 sites while keeping the physical length fixed at $\Lphy=3.4$ fm. By demanding that the ground state is $\Es_L$ for each lattice spacing, we extract the lattice-spacing dependence of the coupling $U(a)$ (see Fig.~\ref{fig:lambda_renorm} left). We also verify using exact diagonalization up to $L= 2048$.

Moreover, taking the infinite length extrapolation ($\Lphy\to\infty$) at fixed lattice spacing yields $U(a)=-4a\gamma^{(s)}_\infty$. We can then derive a relation between the finite-length binding momentum and its infinite-length limit in the form of a transcendental equation,
\begin{align}
\gamma^{(s)}=&\gamma^{(s)}_L\tanh\left(\frac{\gamma^{(s)}_L\Lphy}{2}\right).
  \label{eq:E2Body_continuum_swave}
\end{align}
Since this relationship is independent of the lattice spacing, it must hold in the continuum limit as well. We can drop the $\infty$ subscript on the left-hand side of the above equation. Using the fact that $\epsilon(a)\approx -1/2\mphy a^2$ for small $a$, it is also straightforward to show that 
\begin{align}
\lim_{a\to0} a\epsilon(a) U(a)=&\frac{-2\gamma_L^{(s)}}{\mphy}\tanh\left(\frac{\gamma_L^{(s)}\Lphy}{2}\right)=\frac{-2\gamma^{(s)}}{\mphy},
  \label{eq:Us_continuum}
\end{align}
As we already know, the one-dimensional attractive $\delta$-potential has a unique bound state with the binding energy $\Es=-\mphy U_s^2/4$ \cite{Griffiths_Schroeter_2018}. Therefore, the lattice coupling approaches the continuum coupling in the continuum limit, i.e., $\lim_{a\to0} a \epsilon(a) U(a)=U_s$. Although this expression has a different form, it is consistent with findings reported in Refs.~\cite{Zhu:2019dho,Korber:2019cuq} using L\"{u}sher's formalism.  In particular, Ref.~\cite{Korber:2019cuq} derived the L\"uscher's quantization condition which relates the physical 1D scattering length, $a_1$, to the energy spectrum of the system in a finite box of length $ L_{\rm phy}$. 

Furthermore, we can show that once $U(a)$ is fixed by the ground state of the two-body system in a finite box, it is sufficient to determine its excited states' energies as $a\to 0$ (\cref{fig:lambda_renorm}).

\begin{figure}[h]
    \centering
    \includegraphics[width=\linewidth]{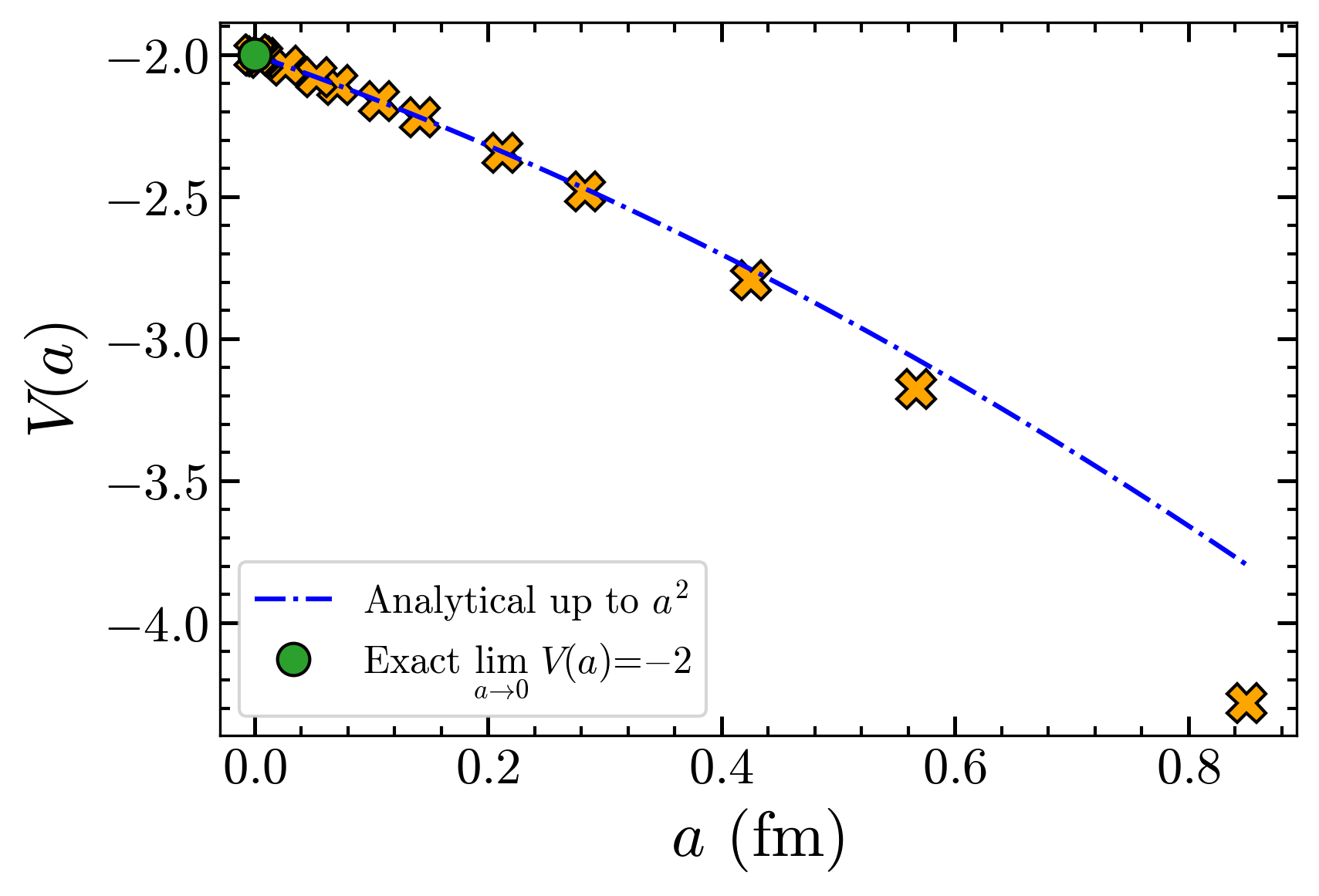}
    \caption{The plot of $V(a)$ as function of lattice spacing. The lattice sizes used in the calculations is up to $L=8192$. The dashed lines are given
by \cref{eq:Vlatt_continuum}.}
    \label{fig:Va_plot}
\end{figure}

\subsection{\texorpdfstring{Two-body $p$-wave channel}{Two-body p-wave channel}}
Similar to the previous section, we can find $\Vlat(a)$ analytically. We introduce a relative momentum $q$ and relative position $\zeta$. A complete set of two particles can be defined as:
\begin{equation}
    \ket{q, \zeta} = \frac{1}{\sqrt{L}}\sum_{x}\psi_{\uparrow,x}^\dagger\psi_{\uparrow,x+\zeta}^\dagger e^{i(2\pi/L)qx}\ket{0}.
\label{eq:1B_basis_pwave}
\end{equation}
We suppress the spin index because they are identical fermions. The Pauli exclusion principle forbids them to occupy the same site, i.e., $\zeta\in\{1,2,...,\lfloor L/2 \rfloor\}$. Note that $\zeta =L/2$ requires  special attention to the normalization factor. It is straightforward to show that $\langle q,L/2|q,L/2\rangle=(1-(-1)^q)$. For even $q$ this state is simply zero. On the other hand, an extra factor of $1/\sqrt{2}$ is required for odd $q$. The $p$-wave interaction term action also does not change the $|q\rangle$ of the state,
\begin{align}
    H_p\ket{q,\zeta}= \epsilon(a) \Vlat(a)\ \delta_{\zeta,1}\ket{q,\zeta},
\end{align}
The expansion the the Green function matrix $G_p(E,a)=(E-H_0-H_p)^{-1}$ reads
\begin{align}
G_p&(E,a)=G_0(E,a)\nonumber\\
&+2\epsilon(a)\ \delta_{\zeta,1}\ \mathcal{G}_p(E,a) \ G_0(E,a)|\zeta\rangle\langle\zeta|G_0(E,a),
\end{align}
where $\mathcal{G}_p=[2/\Vlat(a)-I_{p}(E,a)]^{-1}$. The lattice interaction parameter $ \Vlat (a)$ corresponds to $ 2/I_{p} $,
where  $I_{p}$  is given by
\begin{equation} \label{eq:B32+p}
    I_{p}(E,a) =\frac{4}{L} \sum_{n=1} ^{L/2-1}\frac{\sin^2(2\pi n/L)}{E/2\epsilon(a)-2(1-\cos(2\pi n/L))}.
\end{equation}
Assume $E=\Ep_{L}$ is negative, we can use the Poisson summation formula again (see Appendix \ref{app:Ilatp}) to show that for $a\ll 1$ while $La=\Lphy$ is fixed,
\begin{equation}
\begin{aligned}
    I_p(a) = &~-1+a\gamma_L^{(p)}\coth\left(\frac{\gamma_L^{(p)}\Lphy}{2}\right) \\
    &\qquad-\frac{(\gamma_L^{(p)}a)^2}{2} +\mathcal{O}(a^3),
\end{aligned}    
\end{equation}
where $\gamma_L^{(p)}=\sqrt{\mphy|\Ep_{L}|}$ is the binding momentum defined from the finite-length energy. Clearly, for small $a$, $\Vlat(a)$ has an analytical form given by
\begin{align}
  \Vlat(a)=&-2-2a\gamma_L^{(p)}\coth\left(\frac{\gamma_L^{(p)}\Lphy}{2}\right)\nonumber\\
  &+(a\gamma_L^{(p)})^2\left[1 -2\coth^2\left(\frac{\gamma_L^{(p)}\Lphy}{2}\right)\right].
  \label{eq:Vlatt_continuum}
\end{align}
Therefore,  $\Vlat(a)$ has a well-defined limit as $a\to 0$ independent of the energy (see \cref{fig:Va_plot}). $\Vlat=-2$ is the fixed point of a renormalization group flow in the two-particle channel. The impact of the physics scale is encoded in the higher-order term.  

In the limit $\Lphy\to \infty$ corresponding to the infinite physical length limit, 
\begin{equation}
    -2a\gamma^{(p)}_{\infty}=\Vlat(a)+2.
    \label{eq:gammap_p_to_V}
\end{equation}
Substituting \cref{eq:gammap_p_to_V}  back to \cref{eq:Vlatt_continuum} yields
\begin{equation}
\gamma^{(p)}=\gamma_L^{(p)}\coth\left(\frac{\gamma_L^{(p)}\Lphy}{2}\right).
\label{eq:infinite_energy_relation}
\end{equation}
Since the above relationship is independent of the lattice spacing, $\gamma^{(p)}$ is a genuine continuum quantity. We drop the $\infty$ subscript.  Interestingly, unlike \cref{eq:E2Body_continuum_swave}, it is impossible to tune $\gamma^{(p)}_L$ to reach an arbitrary shallow bound state of the two-body system. Evidently, even if $\gamma_L^{(p)}\to 0$, the binding momentum in the infinite-length limit is bound below at $\gamma^{(p)}\geq 2/\Lphy$, equivalently, $|\Ep|\geq 4/\mphy\Lphy^2$. Note that this is a pure finite-length artifact. The threshold is recovered in a true infinite-length limit.

\section{Few-body results in infinite length and continuum extrapolations } \label{sec:Infinite Volume Extrapolation}
Another important limit of a lattice model is the infinite-length limit \cite{Montvay_Münster_1994}. At the infinite-length limit, the discretized lattice result is supposed to predict the outcome of an experimental measurement \cite{Lenci2010}, or a continuous theory. However, this is not guaranteed for every model, as discussed in Ref.~\cite{PhysRevC.110.024002}. An example would be the finite-system description of the superconductivity theory. Though it successfully reproduces the spectra of low-energy states of many nuclei, this theory fails to approach the normal BCS theory as the system size becomes infinite \cite{PhysRev.135.A1172}. 

Therefore, we want to test our lattice model at the $\Lphy\to\infty$ limit. To approach the limit, the lattice spacing $a$ and its couplings $U(a)$ and $V(a)$ are fixed, and the lattice sites $L$ are increased so that the physical box length increases. At each $L$, we diagonalize the Hamiltonian to obtain the lattice energy, which now can change with \Lphy. It is important to point out that the main source of error comes from the discretization artifact. To obtain the continuum value, we can repeat the calculation for several spacings and extrapolate $a\to 0$. 

This is a distinct procedure from the tuning of the couplings discussed in \cref{sec:fixing_2B_couplings}. Previously, we fix $\Lphy$ and $\Es_L$ and $\Ep_L$ and solve for $U(a)$ and $V(a)$ as $a$ varies. In this section, we will choose $U(a)$ and $V(a)$ for a given spacing based on \cref{sec:fixing_2B_couplings} and then increase the number of sites to study how the few-body systems' ground state energies depend on the box size. Taking the infinite-length limit at fixed lattice spacing, followed by the continuum limit, we show that the lattice can reproduce the exact continuum physics of few-body systems. 

\subsection{\texorpdfstring{$s$}{s}-wave interaction only \label{sec:swave_fewbody}}
\begin{table}[h]
    \centering
    \begin{tabular}{l c c r}\hline
        $a$ [fm] &$L$& $U(a)$& $\epsilon(a)$ [MeV]\\\hline
         $0.2125$&$1024$&$ -0.28270884 $& $458.5427$\\
          $0.10625$ &$2048$&$-0.14237550$& $1834.1666$\\
         $0.053125$ & $2048$ &$-0.07131545$&$7336.6665$  \\\hline
    \end{tabular}
    \caption{Values of $\epsilon(a)$ and $U(a)$ at three lattice spacings. They are obtained by using the toy parameters
given in \cref{tab:input}. The $L$ column gives the number of sites used for the infinite-length extrapolation in both calculations. }
    \label{tab:lattice_parameters_swave}
\end{table}
\subsubsection{\texorpdfstring{Two-body system $(\uparrow\!\downarrow)$}{Two-body system (up-down)}}
We first verify that the lattice model reproduces the known continuum two-body physics. We have already shown earlier that $U_s=\lim_{a\to0}a \epsilon(a)U(a)$, which is $-0.140966$ using the toy parameters from \cref{tab:input}. The exact continuum two-body energy is then given by 

\begin{equation} \label{eq:QMcalculation}
   \Es=-\frac{\mphy U^2_s}{4}=-4.66975~{\rm MeV}.
\end{equation}
We calculate the ground state energy $\Es_L$ via exact diagonalization using the reduced basis introduced in \cref{sec:s_renorm}. All error is systematic, and arises from two extrapolations $\Lphy\to\infty$ and $a\to0$. The finite-length is well controlled because the reduced basis makes the large-$L$ accessible. We can examine the convergence directly without assuming a particular asymptotic functional form. We compare energies at the largest values of $L$ and quote the number of stable significant figures as a conservative bound on the residual. 
We find that the dominant uncertainty due to  $a\to0$ extrapolation is dominant. We estimate the statistical error from the spread of extrapolated values obtained by varying the fit form (including or omitting the next order in $a$) and varying the fit window (dropping the coarsest or the finest spacing). We report the central value from the preferred fit and this spread as the systematic uncertainty. This procedure is used throughout the remainder of the paper.

\begin{figure}[h]
    \centering
    \includegraphics[width=\linewidth]{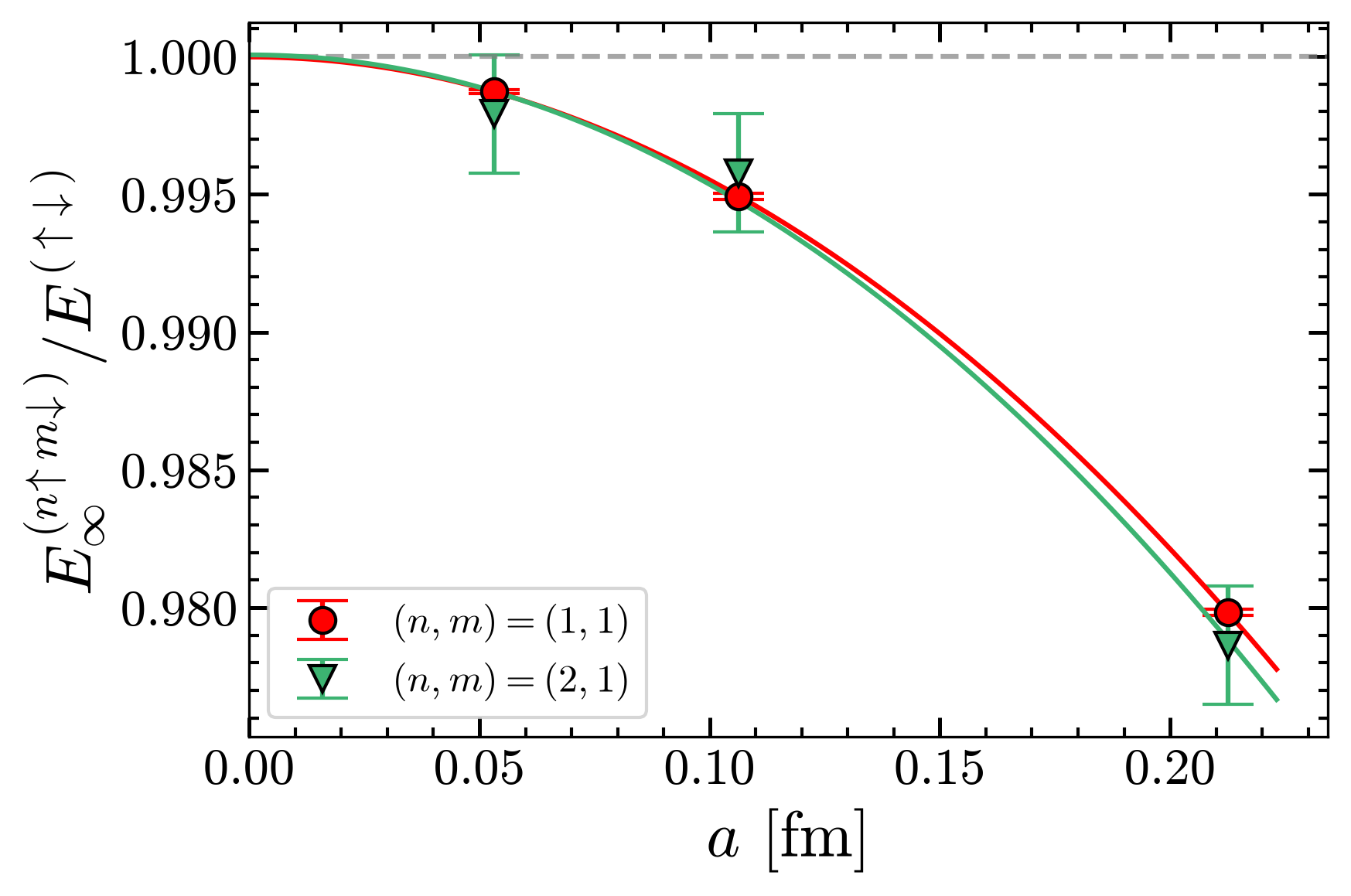}
    \caption{Ratio of the infinite-length two-body (red) and three-body (green) ground-state energies to the continuum two-body ground-state energy, at three lattice spacings. The solid lines are quadratic fits to the data.}
    \label{fig:twoB_error}
\end{figure}

The lattice results exhibits an $\mathcal{O}(a^2)$ scaling (see \cref{fig:twoB_error}). In the dual infinite-length and continuum limit, we find that $\Es_\infty=-4.6696(3)~{\rm MeV}$. The two-body sector's ground state energy is reproduced with high accuracy. This certifies that once the coupling is correctly renormalized, the lattice reproduces the continuum infinite-length physics.

\subsubsection{Three-body system $(2\!\uparrow+1\!\downarrow)$}
It is straightforward to extend this calculation to a three-body system consisting of two $\psi_\uparrow$ particles and one $\psi_\downarrow$ particle.  We define the complete set of three-particle states
\begin{equation}
|\bq, \delta_1, \delta_2\rangle = \frac{1}{\sqrt{L}}\sum_{x} \psi^\dagger_{\uparrow,x+\delta_1}\psi^\dagger_{\uparrow,x+\delta_2}\psi^\dagger_{\downarrow,x}e^{\frac{2\pi i}{L}\bq x}|0\rangle,
\end{equation}
where $0\leq\delta_1<\delta_2\leq L-1$. It is straightforward to show that $\langle\bq’, \delta’1, \delta’2|\bq, \delta_1, \delta_2\rangle=\delta{qq’}\delta{\delta_1’\delta_1}\delta_{\delta_2’\delta_2}$. Due to the action of the $H_0$ terms, the two $\psi_\uparrow$ particles can change sites such that $\delta_1>\delta_2$. In this case, we need to reorder and include a minus sign. Furthermore, we can show that
\begin{align}
H_s|\bq,\delta_1,\delta_2\rangle =\epsilon(a) U(a)\delta_{r_1,0}|\bq,\delta_1,\delta_2\rangle .
\end{align}
Again, the Hamiltonian is block diagonal in $|q\rangle$. For the $2\uparrow+1\downarrow$ system, the three-body ground state is double degenerate and exists in the $q=\pm1$ sector. The result is presented in \cref{fig:twoB_error}. As we can see from the figure, the error due to infinite-length extrapolation is still quite large despite calculating at large $\Lphy$. Because the three-body bound state is very close to the two-body, it implies that there is no genuine three-body bound state in the $\uparrow\uparrow\downarrow$. Physically, the
ground state of $2\uparrow+1\downarrow$ is a single $\uparrow\downarrow$
dimer with the remaining free $\uparrow$ fermion acting as a spectator. The residual energy above the two-body threshold is therefore proportional to $1/\Lphy^2$. This power-law approach is in sharp contrast to the exponential $\exp[-\gamma^{(s)} \Lphy]$ suppression that governs a genuine bound state \cite{Luscher:1985dn,Luscher:1986pf}, which accounts for slow convergence to the infinite-length limit.

If we assume that the $\mathcal{O}(a^2)$ scaling holds for a three-body system, in the continuum limit, $ \Ethrees_{\infty}/\Es=-1.000(2)$ MeV.  The absence of a new trimer state is a well-known consequence of the
Bethe-ansatz solution of the one-dimensional two-component Fermi gas~\cite{Yang:1967bm,GAUDIN196755}. The structure of its bound states follows from the classification of string solutions of the Bethe
equations~\cite{takahashi1971one}.

We also try to extend the exact diagonalization method to study the ground state energies of the
$3\!\uparrow+1\!\downarrow$ and $2\!\uparrow+2\!\downarrow$ systems. They are converging  with increasing $\Lphy$, but have not yet reached the asymptotic infinite length regime due to the lack of binding energy relative the two-body threshold. In particular, $3\!\uparrow+1\!\downarrow$ configuration is a special case of a highly polarized Fermi gas with a single impurity, its exact continuum ground state energy approaches $\Es$ \cite{10.1063/1.1704798,Castella_1993}.

\subsection{\texorpdfstring{$p$}{p}-wave interaction only}
Reference~\cite{McGuire_1964} has shown that in the one-dimensional $N$-body bosonic system interacting via an attractive pairwise $\delta$ potential, the two-body binding energy is the sole scale and the ratio between the $N$-body bound state and $\Ep$ is given by \cite{McGuire_1964} 
\begin{equation}
    \frac{E^{(N\uparrow)}}{\Ep}=\frac{N(N^2-1)}{6},
    \label{eq:Mcguire}
\end{equation}
which is independent of the strength of the potential.  This result carries over to $N$-body spinless (or fully spin-polarized) fermions interacting via attractive $p$-wave contact interaction through the Cheon–Shigehara duality \cite{Cheon:1998iy,Cheon_1998}. Setting $U=0$, the lattice theory studied in this work (\cref{eq:c_hamiltonian}) is a discretization of the $p$-wave fermion model. In the following sections, we show that our lattice construction reaches $E^{(N\uparrow)}/\Ep=4,10,$ and $20$ for $N=3,4$, and $5$, respectively, in the continuum limit. 

\subsubsection{Two-body system $(\uparrow\!\uparrow)$}
\begin{figure}[h]
    \centering
     \includegraphics[width=.95\linewidth]{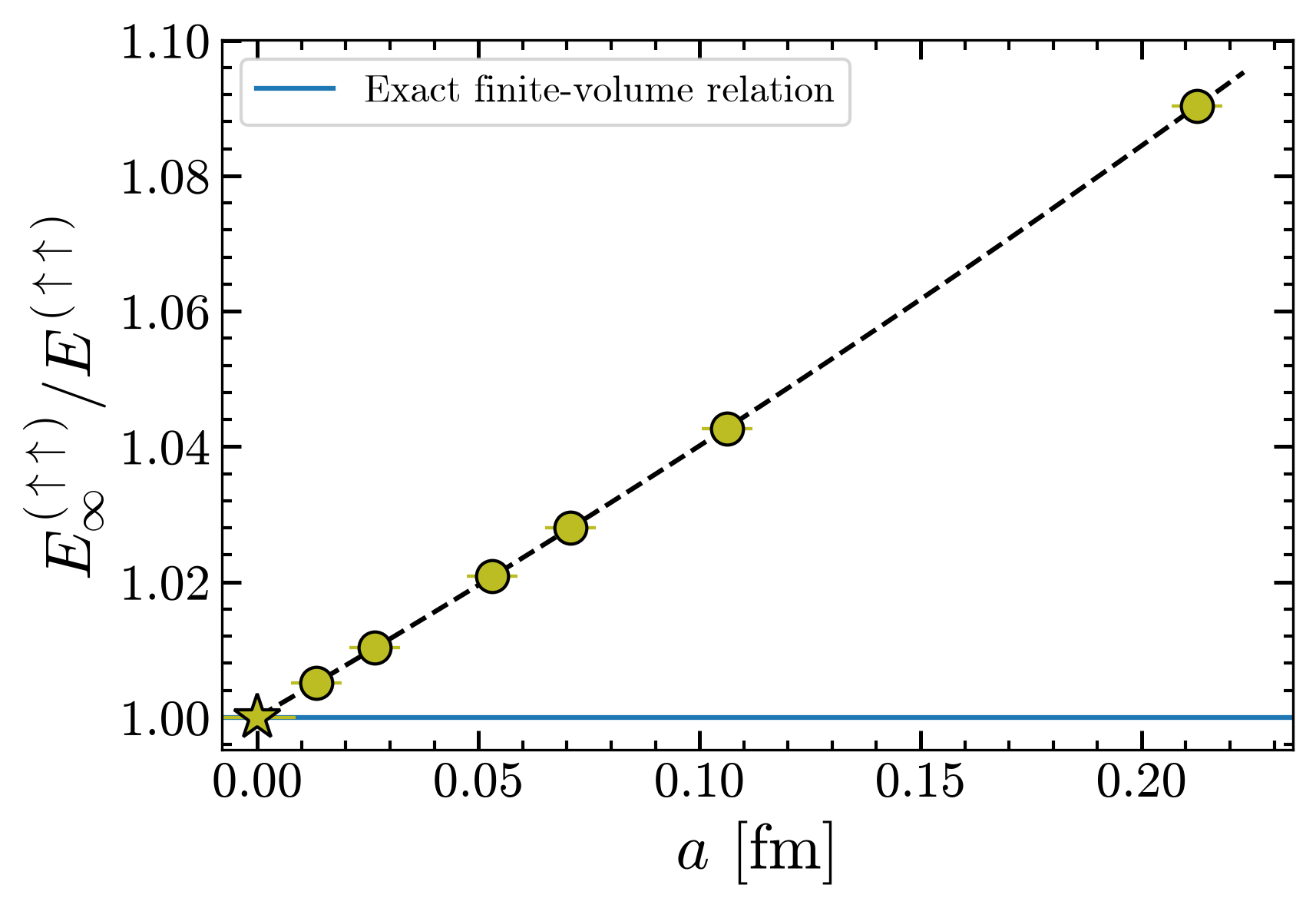}
    \caption{Lattice-spacing dependence of $\Ep$ at $\Lphy\to \infty$. As the lattice spacing becomes smaller, the number of lattice sites is increased such that $La=\Lphy=108.8$ fm. The solid line is a linear fit to the data.}
    \label{fig:2B_binding}
\end{figure}

We start first with a trivial test on the two-body system. Using the toy parameters from \cref{tab:input}, \cref{eq:infinite_energy_relation} yields $\Ep= -21.4144$ MeV in the continuum. As before, we will use the coupling $V(a)$ at several lattice spacings (see \cref{tab:p-wave_infinite_length}).  \cref{fig:2B_binding} shows a linear fit in $a$ gives $\Ep_\infty= -21.4146(3)$ MeV.

\begin{table}[h]
    \centering
    \begin{tabular}{c c}\hline
        $a$ [fm] & $V(a)$\\\hline
        $0.2125$&  $-2.34558958 $ \\
          $0.10625$ & $-2.16224326$\\
       $0.070833$ & $-2.10599051$ \\
         $0.053125$ & $-2.07870036$ \\
          $0.0265625$ & $-2.03876877$ \\
          $0.01328125$ & $-2.01924184$ \\\hline
    \end{tabular}
    \caption{Values of $V(a)$ at several lattice spacings. They are obtained by using the toy parameters
given in \cref{tab:input}.}
    \label{tab:p-wave_infinite_length}
\end{table}
\subsubsection{Three-body system $(3\uparrow)$\label{sec:3polarized}}
In this section, we calculate the ground state energy of the three-$\psi_\uparrow$ system. For exact diagonalization, we define the momentum-projected relative-coordinate state as 
\begin{equation}
    |q,\zeta_1,\zeta_2\rangle=\frac{1}{\sqrt{L}}\sum_{x}\psi^{\dagger}_{\uparrow,x}\psi^{\dagger}_{\uparrow,x+\zeta_1}\psi^{\dagger}_{\uparrow,x+\zeta_2}e^{-i2\pi qx/L}|0\rangle,
\end{equation}
where $0<\zeta_1<\zeta_2<L$. Notice that this basis is three-fold over-complete because these fermions are identical. The three representations of the same state are $|q,\zeta_1,\zeta_2\rangle$, $|q,\zeta_2-\zeta_1,L-\zeta_1\rangle= e^{-i2\pi \zeta_1 q/L }|q,\zeta_1,\zeta_2\rangle$, and $|q,L-\zeta_2,L-\zeta_2+\zeta_1\rangle = e^{-i2\pi\zeta_2 q/L}|q,\zeta_1,\zeta_2\rangle$. It is possible to remove this redundancy by defining the canonical state to be the lexicographically smallest among these three. In particular, a hop that lands on a non-canonical
label is mapped back to its canonical partner, carrying the additional phase and sign. If the number of lattice sites is divisible by 3, the state $|q,L/3,2L/3\rangle$ only exits in sectors with $q\equiv 0~(\rm mod~ 3)$. The normalization factor is  $1/\sqrt{3L}$ rather than $1/\sqrt{L}$. 
\begin{align}
    H_p|\bq,\zeta_1,\zeta_2\rangle &= \epsilon(a) V(a)\big(\delta_{\zeta_1,1}\nonumber\\
    &~+\delta_{\zeta_2,L-1} +\delta_{\zeta_2-\zeta_1,1}\big)|\bq,\zeta_1,\zeta_2\rangle .
\end{align}
For three identical fermions, the ground state exists in $q=0$ sector. We confirm by checking all momentum sectors at small $L$. The result is presented in \cref{fig:2B_binding}. Unlike the $\uparrow\!\uparrow\!\downarrow$ system discussed above, the $\uparrow\!\uparrow\!\uparrow$ is deeply bound. The convergence to $\Lphy\to\infty$ is fast. The continuum extrapolation yields $\Ethreep_\infty/\Ep_\infty=3.999(1)$, which is identical to that of a three-boson bound state found by McGuire \cite{McGuire_1964}. The lattice model with a nearest-neighbor interaction thus correctly reproduces the continuum result, with the two-body coupling $V(a)$ as the only input.
\begin{figure}[h]
    \centering
     \includegraphics[width=\linewidth]{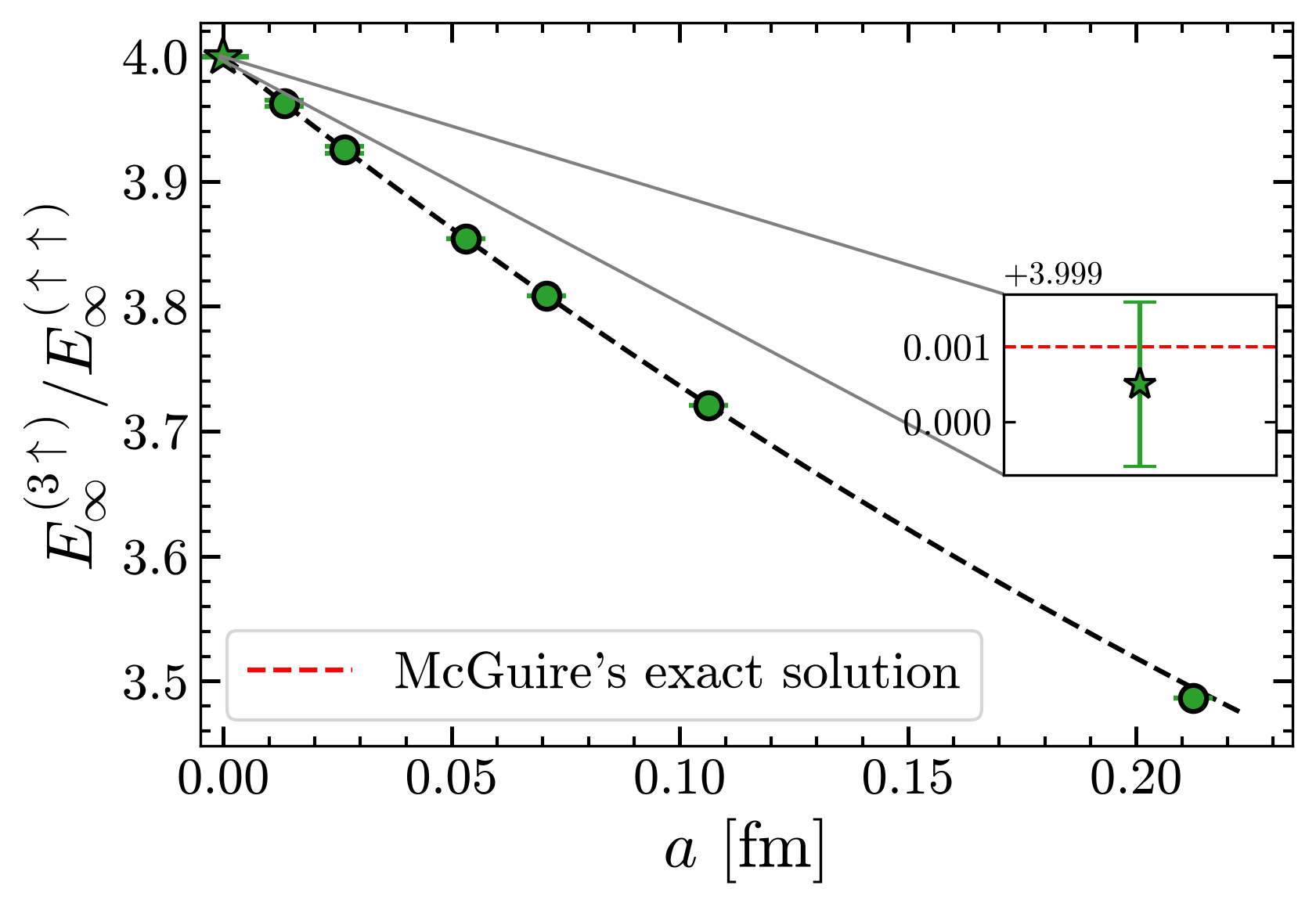}
    \caption{The plot of the ratio $\Ethreep_\infty/\Ep_\infty$  as a function of the lattice spacing. The dashed black line is a quadratic fit to data. }
    \label{fig:3B_binding}
\end{figure}

\subsubsection*{Improving convergence to the continuum limit with a three-body force}
Our previous section demonstrates that no three-body term is required to reproduce the three-body ground state energy. On the lattice, the Pauli principle forbids identical fermions from occupying the same site, so the three-body contact term familiar from the bosonic theory or field theoretical approach (see, e.g., Ref.~\cite{PhysRevA.103.043307}) is absent. The minimum  three-body operator can take the form of 
\begin{equation}
    H_{\rm 3} = \epsilon(a) h(a) \sum_{i} n_{i-1} n_i n_{i+1}.
\end{equation}
This term is not a counterterm required by renormalization.
But it can serve as a lattice improvement operator in the spirit of Ref.~\cite{Klein:2018iqa}, which implements Symanzik’s improvement program to remove systematic
lattice errors for a one-dimensional system of bosons with zero-range interactions. The three-body operator is not the only improvement term. For example, Ref.~\cite{Klein:2018iqa} also includes a squared-momentum-transfer term, a Galilean-invariance-restoring term, and an improved kinetic term.

In the reduced basis, the three-body interaction can be written as
\begin{align}
    H_3|\bq,\zeta_1,\zeta_2\rangle &= \epsilon(a) h(a)\big(\delta_{\zeta_1,1}\delta_{\zeta_2,2}+\delta_{\zeta_1,1}\delta_{\zeta_2,L-1}\nonumber\\
    &~ +\delta_{\zeta_1,L-2}\delta_{\zeta_2,L-1}\big)|\bq,\zeta_1,\zeta_2\rangle .
\end{align}
For each fixed lattice spacing $a$, we calculate the dimensionless three-body interaction strength $h(a)$ such that at the infinite-length limit, $E^{\mathrm{(3B)}}_\infty/E^{\mathrm{(2B)}}_\infty=4$. With the data of $h(a)$, we can infer the dimensionless three-body interaction strength at the continuum limit by performing a fitting function.
\begin{figure}[h]
    \centering
    \includegraphics[width=\linewidth]{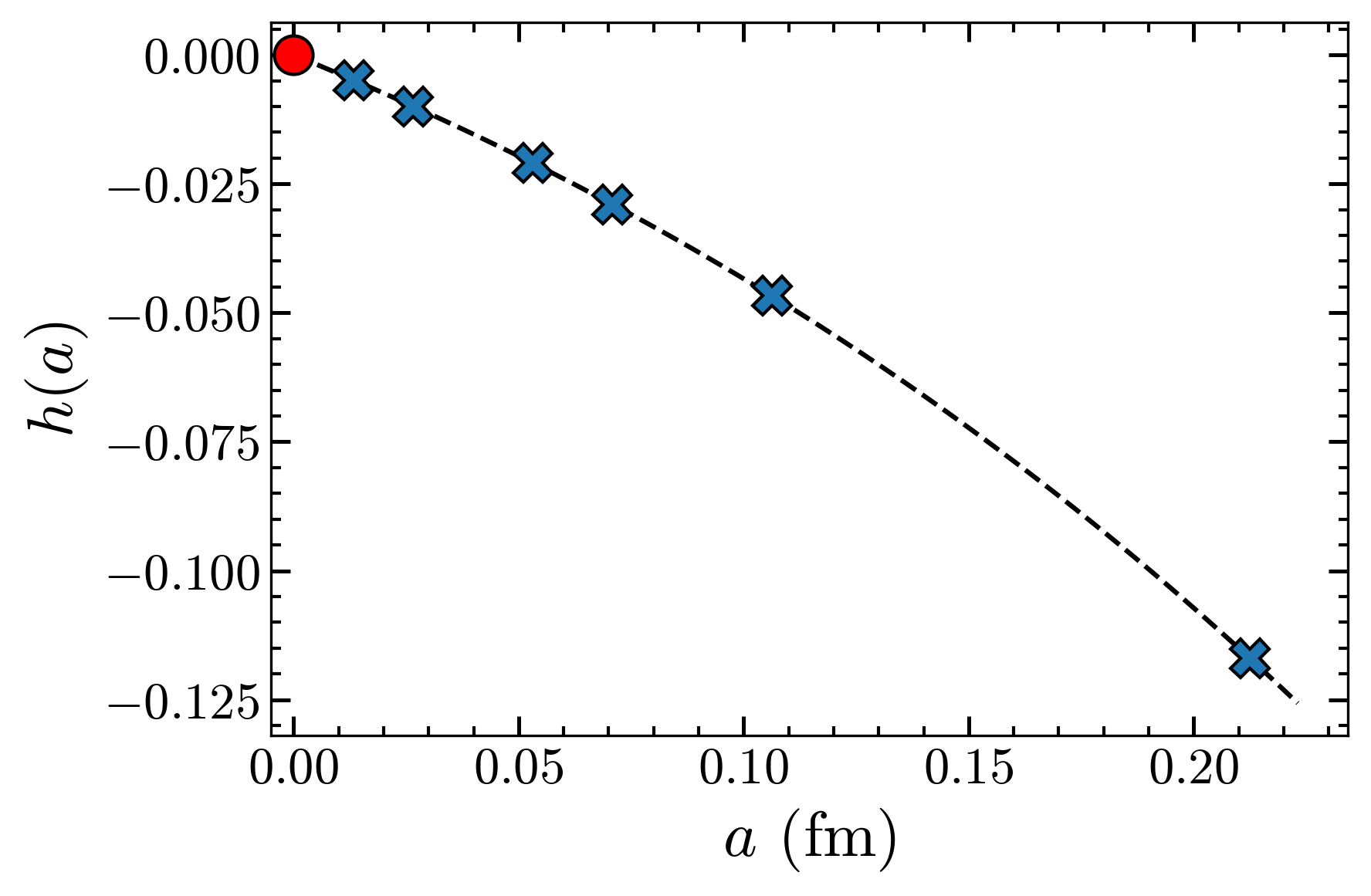}
    \caption{ The plot of three-body lattice coupling $h(a)$ as a function of the lattice spacing such that $\Ethreep_\infty/\Ep_\infty = 4$. The dashed line represents  a quadratic fit.}
\label{fig:Infinite_volume_three_body_force}
\end{figure}
The fitting shows that, at the continuum limit, $h(a)$ is consistent to 0 within the extrapolation error.

\begin{figure*}[t]
    \centering
    \includegraphics[width=0.46\linewidth]{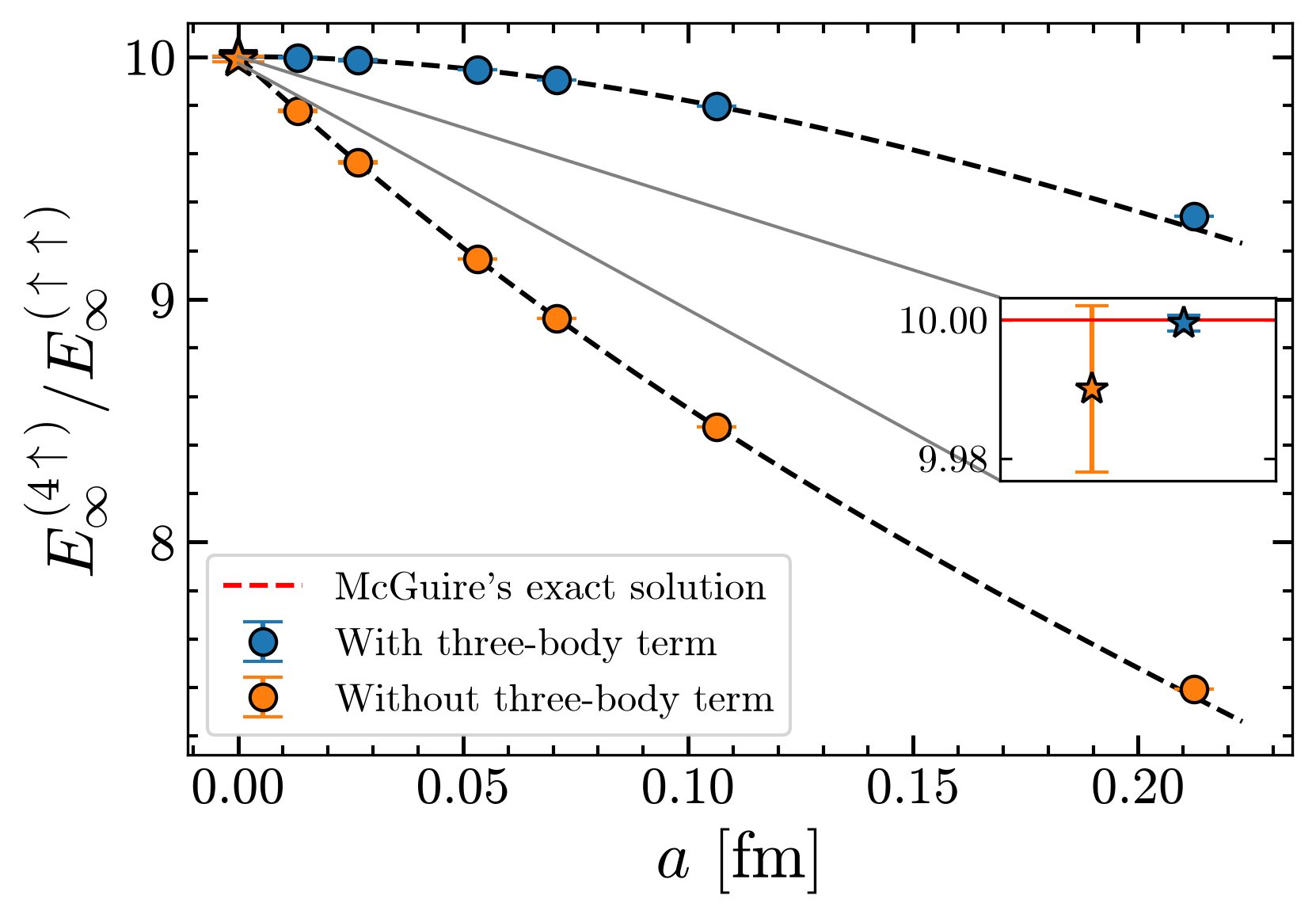}
    \includegraphics[width=0.47\linewidth]{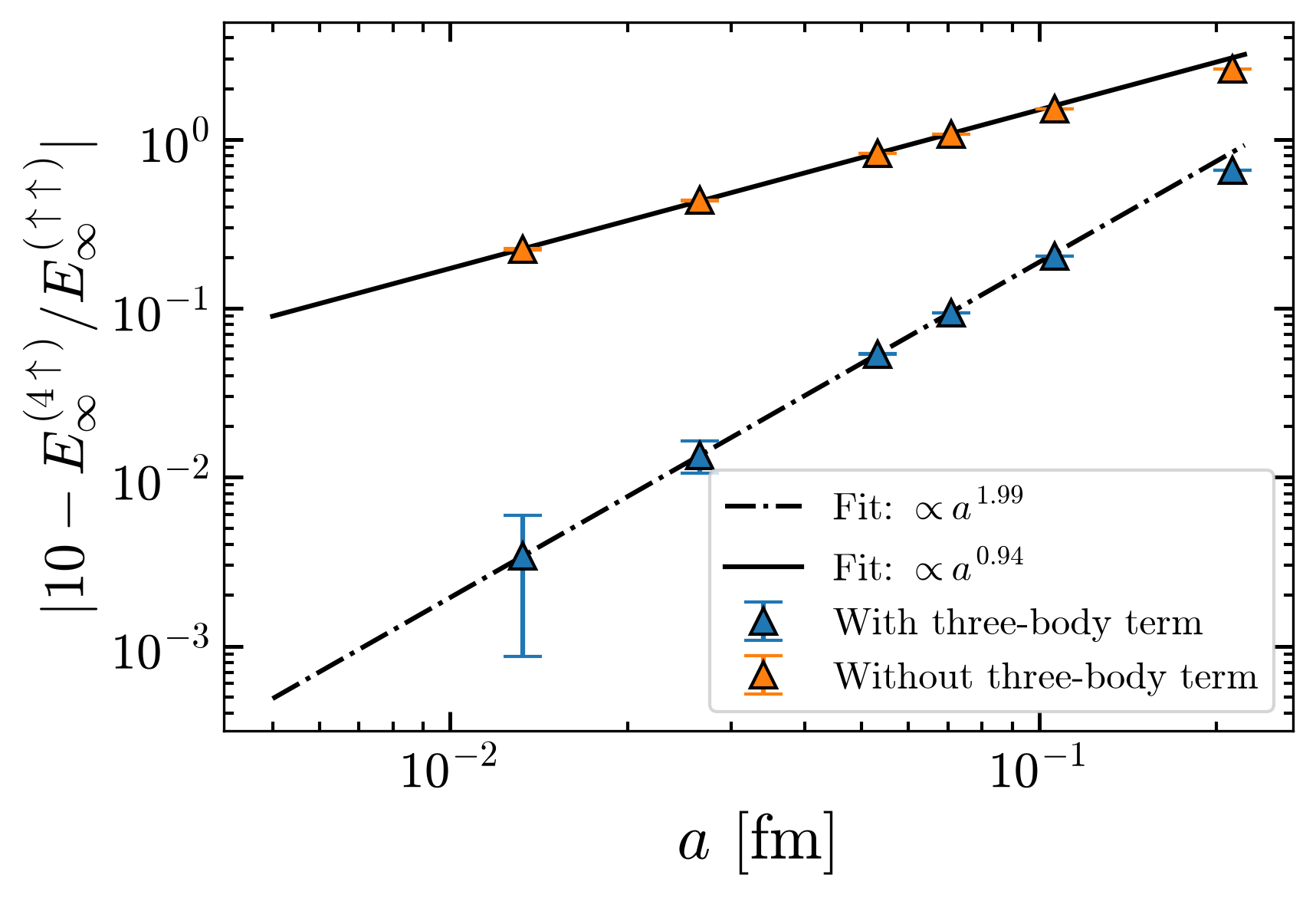}
    \caption{ The plot of the ratio $\Efourp/\Ep$ (left) and the deviation from the benchmark value of 10 (right) at different lattice spacings. For the benchmark plot, two coarsest points are not included in the fittings.}
    \label{fig:Infinite_volume_four_body_force}
\end{figure*}
\subsubsection{Four- and five-body systems \label{sec:four+five_polarized}}

\begin{figure}[h]
    \centering
    \includegraphics[width=\linewidth]{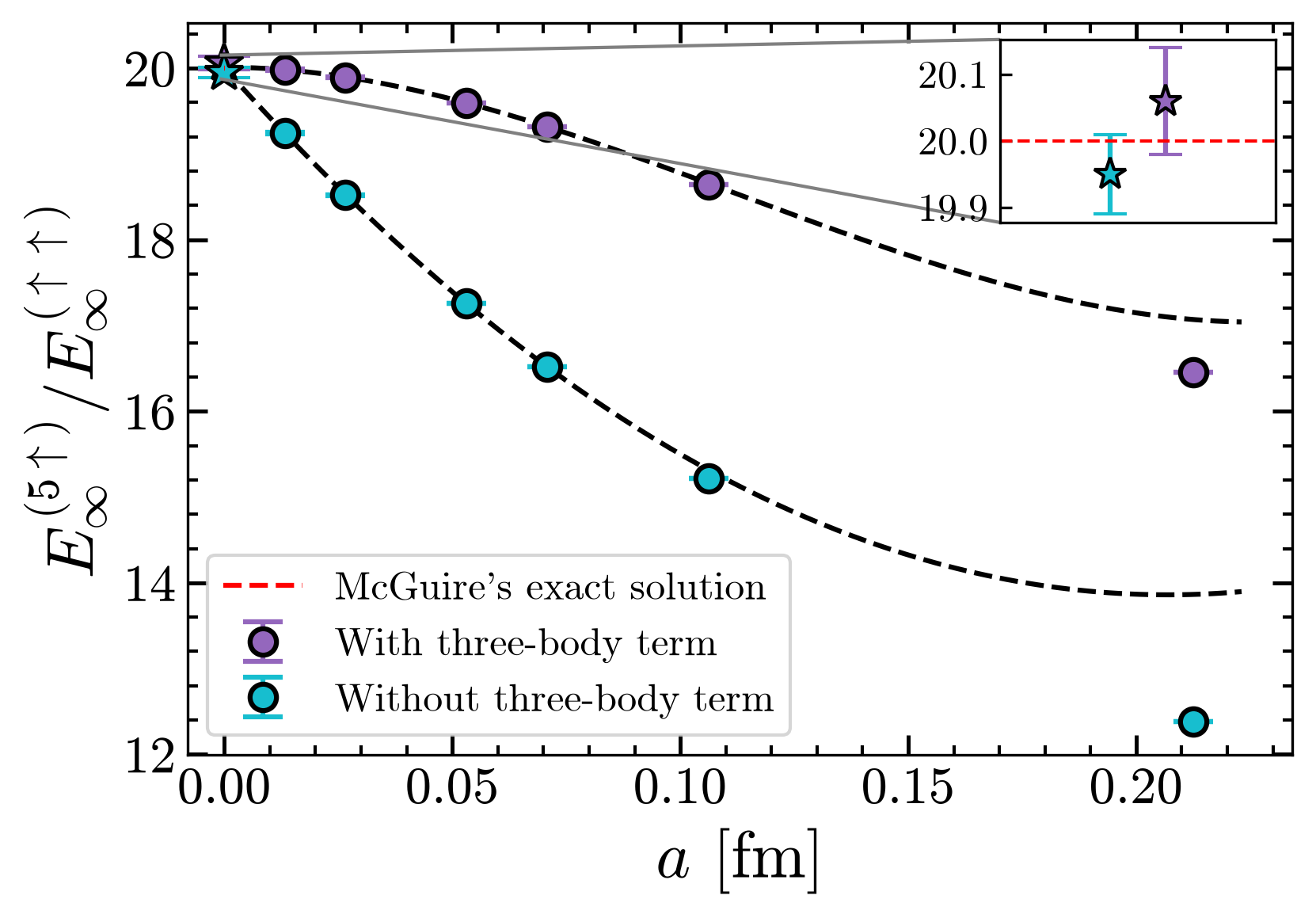}
    \caption{The plot of the ratio $\Efivep/\Ep$ as function of the lattice spacing.}
    \label{fig:Infinite_volume_five_body_force}
\end{figure}

We extend the analysis to the four- (\cref{fig:Infinite_volume_four_body_force} left) and five-body (\cref{fig:Infinite_volume_five_body_force})  sectors of identical fermions, in each case computing the spectrum both with and without the three-body term. We obtain $\Efourp/\Ep$ extrapolates to $9.99(1)$ without the three-body term and $10.000(1)$ with it, against the exact value of 10 and  $\Efivep/\Ep$ extrapolates to $19.95(6)$ without the three-body term and $20.06(8)$ with it, against the exact value of 20. They all agree with the benchmark continuum value within the uncertainties. Thus, the correct few-body physics is reproduced from the two-body coupling alone. 

The three-body term, as expected, improves the rate of convergence. For illustration, we examine the deviation of the calculated four-body ratio from the exact value and find that the convergence is $\mathcal{O}(a)$ without the three-body term and  becomes $\mathcal{O}(a^2)$ when it is included (see \cref{fig:Infinite_volume_four_body_force} right). Removing the leading linear lattice artifact improves the accuracy at the finest spacing by roughly two orders of magnitude and tightens the extrapolation uncertainty by an order of magnitude. 

\section{Three- and Four-body systems with coexistent \texorpdfstring{$s$}{s}- and \texorpdfstring{$p$}{p}-wave interactions}
\label{sec:universality}

\subsection{\texorpdfstring{Three-body system $(2\!\uparrow+1\!\downarrow)$}{Three-body system (up-up-down)} }
In this section, we will revisit the system consisting of two $\psi_\uparrow$ and one $\psi_\downarrow$ fermions, where both $s$- and $p$-wave interactions present. Since they acts on different pairs, As discussed in the previous section, the only scales that survive the dual infinite-length and continuum limits are the two-body binding energies, $\Es$ and $\Ep$.  Writing the trimer energy as $\Ethrees$ and the lowest two-body
breakup threshold as
\begin{equation}
  E_{\rm thr}=\min\left[\Es,\,\Ep\right],
\end{equation}
we want to verify whether the dimensionless ratio $\Ethrees/E_{\rm thr}$ can depend only on the combination of the two-body scales,
\begin{equation}
  \mathcal{R}^{(3)}(\eta)=\frac{\Ethrees}{E_{\rm thr}},
  \qquad
  \eta\equiv\frac{\Ep}{\Es} .
  \label{eq:univ}
\end{equation}
If this is the case, any system with the same coupling ratio inherits the same curve and is independent of the microscopic details of the interaction.

\renewcommand{\arraystretch}{1.5}
\setlength{\tabcolsep}{15pt}
\setlength{\arrayrulewidth}{0.2mm}
\begin{table}[htbp]
    \centering
    \begin{tabular}{c c c}
        \hline
        $a$ [fm] & {Set I} & Set II \\
        \hline
        0.21250 & 2.49414176 & 2.57483126 \\
        0.10625 & 2.67507883 & 2.71676052 \\
        0.07083 & 2.73866708 & 2.76622931 \\
        0.05313 & 2.77046030 & 2.79094661 \\
        0.02656 & 2.81764992 & 2.82768436 \\
        0.01328 & 2.84084110 & 2.84578957 \\
        \hline
        $a\to0$ & 2.8646(3) & 2.8643(2)\\
        \hline
    \end{tabular}
        \caption{The ratio $\mathcal{R}^{(3)}(\eta)$ at $\eta=1/4$ for two different choice of two-body binding energy input which is used to fix the lattice couplings $U(a)$ and $V(a)$. Set I, $(\Es,\Ep)=(-85.6576,-21.4144)$ MeV, and Set II, $(\Es,\Ep)=(-61.4284,-15.3571)$ MeV.  }
            \label{tab:eta1/4}

\end{table}

\begin{figure*}[t]
    \centering
    \includegraphics[width=.95\linewidth]{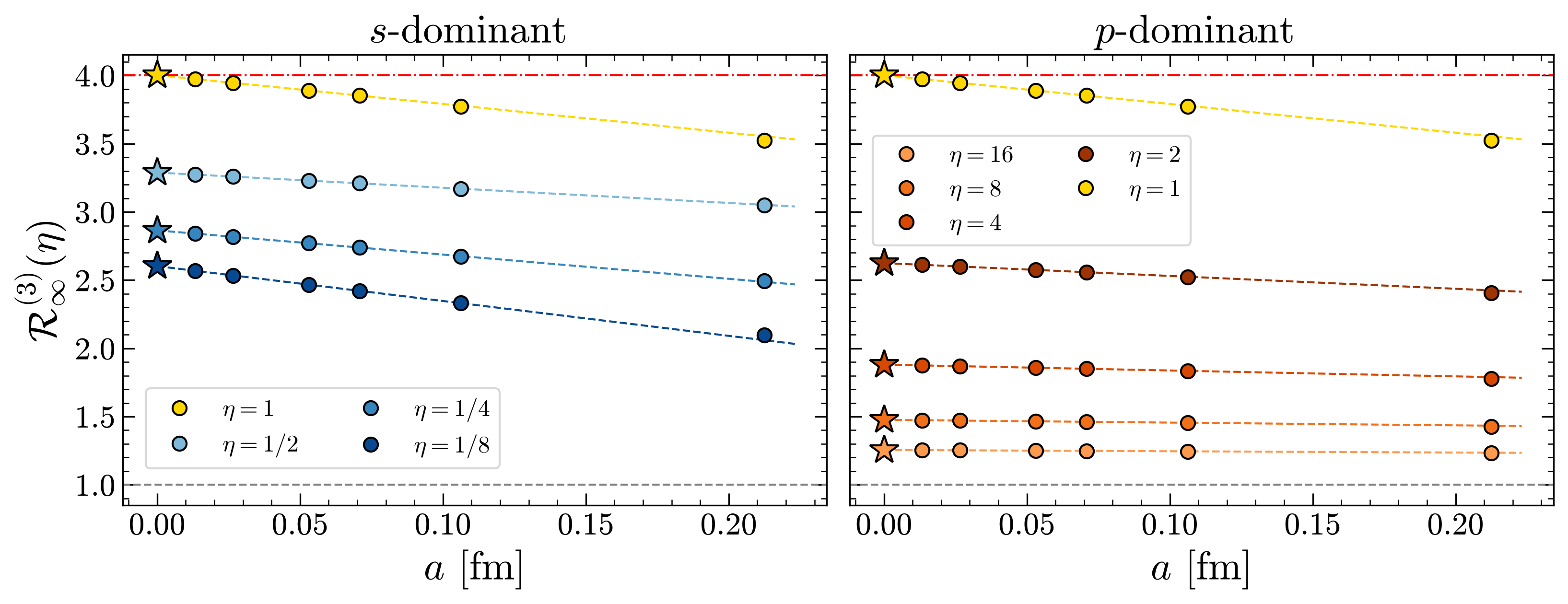}
    \caption{The ratio $\mathcal{R}^{(3)}(\eta)$ as function of lattice spacing, at several $\eta = \Ep/\Es$ values. We separate the results into two domains $|\Es|>|\Ep|$ (left)  and $|\Es|<|\Ep|$ (right). The red dot-dashed line represents the exact ratio $E^{(3\uparrow)}/E^{(\uparrow\!\uparrow)}=4$. }
    \label{fig:sp_converging}
\end{figure*}
We can observe this behavior by calculating $\mathcal{R}^{(3)}$ using different combination of $\Es$ and $\Ep$ such that the value of $\eta$ remains the same. We no longer need to use only the lattice couplings $U(a)$, $V(a)$ previously derived with the toy input parameters. Any two-body binding energies, produced from properly tuned lattice couplings according to \cref{eq:Ulatt_continuum,eq:E2Body_continuum_swave,eq:Vlatt_continuum,eq:infinite_energy_relation}, yielding the same $\eta$ can be used to calculate the corresponding $\mathcal{R}^{(3)}$ values.

In \cref{tab:eta1/4}, we choose a representative point $\eta=1/4$, the extrapolated values $\mathcal{R}_\infty^{(3)}(\eta=1/4)$ are in excellent agreement in both choices. Thus the independence of the ratio on the individual couplings at fixed $\eta$ is confirmed numerically. For the final calculations, we will keep $\Ep$ fixed at $-21.4144$ MeV and vary $\Es$ to achieve the desired $\eta$ value. We then repeat this procedure for each $\eta$ (see \cref{fig:sp_converging}).  

\begin{figure}[h]
    \centering
    \includegraphics[width=\linewidth]{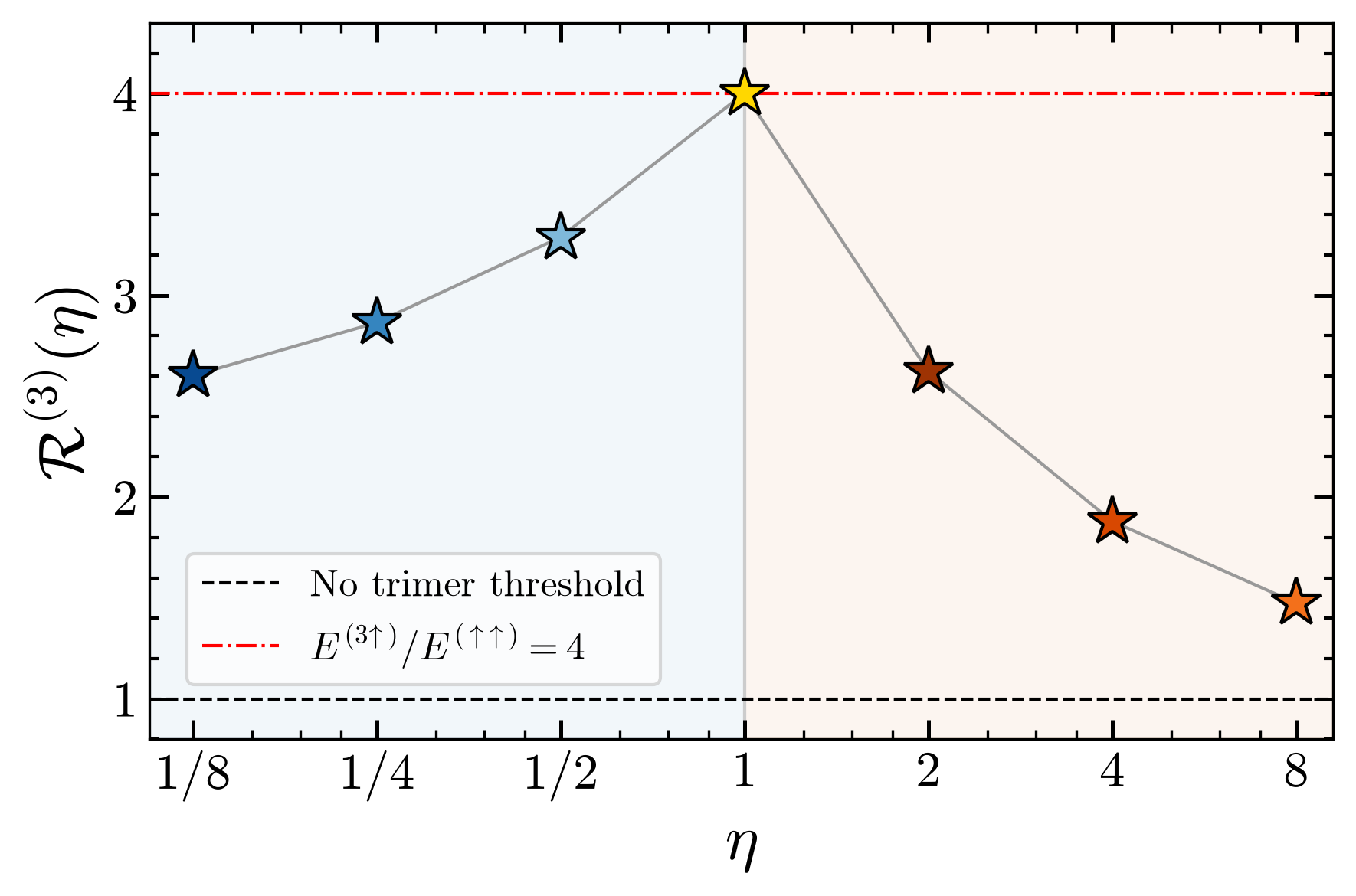}
    \caption{The plot of continuum ratio $\mathcal{R}^{(3)}$ versus $\eta$. }
    \label{fig:3B_phasediagram}
\end{figure}

The dependence of the continuum $\mathcal{R}^{(3)}$ on $\eta$ is shown in \cref{fig:3B_phasediagram}. Noticeably, when
$\eta=1$ the ratio is  $\mathcal{R}^{(3)}=4.001(1)$, which is equal to that of three identical fermions (\cref{sec:3polarized}). We note that the ground state of $2\!\uparrow+1\!\downarrow$ lies in the $q=\pm 1$ sector, whereas that of three identical fermions lies at $q=0$. This is a consequence of Pauli shell filling. The kink at $\eta=1$ is due to $E_{\rm th}$ switches from $\Es$ in $s$-dominant regime to $\Ep$ in $p$-dominant regime. 

Note that the limit $\eta\to0$ cannot be approached directly on a finite lattice by tuning either $|\Es|\to \infty$ or $\Ep\to 0$. A deeply bound $\Es$ produces a compact state that requires an extremely fine lattice spacing to resolve. On the other hand $\Ep$ is bounded below due to the finite-length effect. The point $\eta=0$ is a distinct configuration in which the $p$-wave interaction is simply
switched off. When $V(a)=0$ there is no $p$-wave dimer, the threshold reduces to $E_{\rm thr}=\Es$. We already show that the ratio is simply 1 (see \cref{sec:swave_fewbody}).

\subsection{Four-body system ($3\!\uparrow+1\!\downarrow$)}
In the four-body sector, the coexistence of $s$- and $p$-wave interactions produces additional binding beyond the 
$s$-channel alone. The resulting tetramer state converges faster as $\Lphy\to \infty$ and $a\to 0$. To reach the infinite-length asymptotic, the largest lattice site used in this analysis is $L=448$. We will focus on the $3\!\uparrow+1\!\downarrow$ configuration.  We find that the dimensionless ratio $\Efour/E_{\rm thr}$, which will be denoted as $\mathcal{R}^{(4)}$, also depend  on the ratio $\eta$ of the two two-body energies only, regardless of their individual strength. 

\begin{figure*}[t]
    \centering
    \includegraphics[width=.95\linewidth]{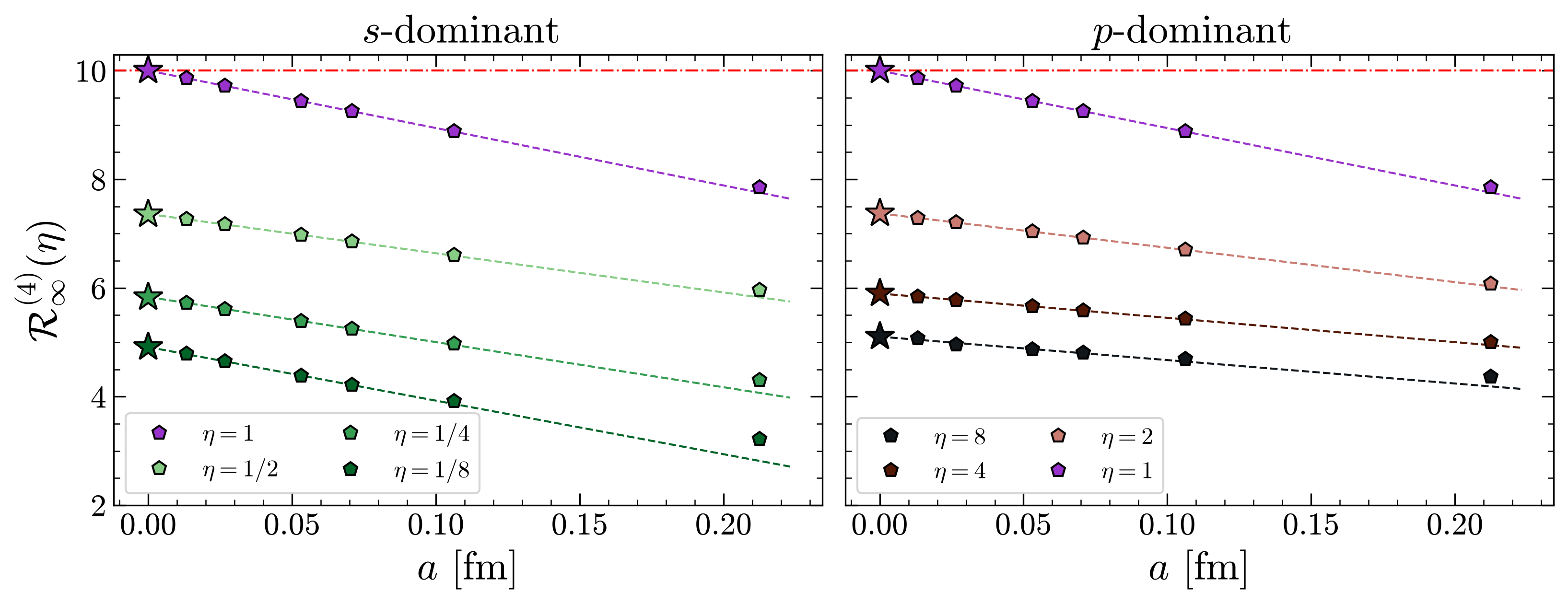}
    \caption{The ratio $\mathcal{R}^{(4)}(\eta)$ as function of lattice spacing, at several $\eta = \Ep/\Es$ values. Again, we separate the results into two domains $|\Es|>|\Ep|$ (left)  and $|\Es|<|\Ep|$ (right). The red dot-dashed line represents the exact ratio $E^{(4\uparrow)}/E^{(\uparrow\!\uparrow)}=10$. }
    \label{fig:sp_converging_31}
\end{figure*}

\begin{figure}[h]
    \centering
    \includegraphics[width=\linewidth]{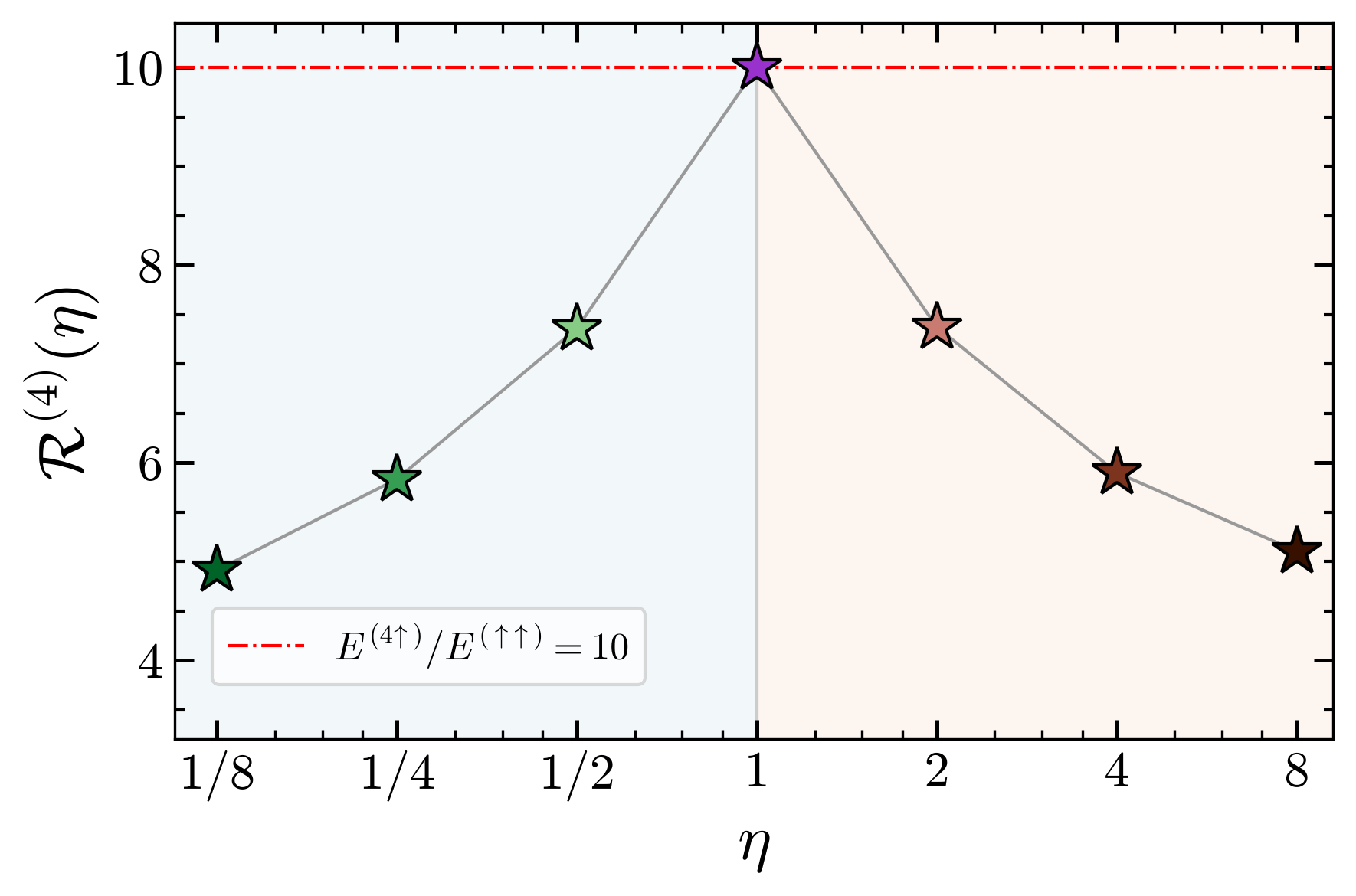}
    \caption{The plot of continuum ratio $\mathcal{R}^{(4)}$ versus $\eta$. }
    \label{fig:4B_phasediagram}
\end{figure}

Following the same procedure and couplings from the previous section, we calculate the infinite-length extrapolated ratio $\mathcal{R}_\infty^{(4)}$ as $a\to 0$ (see \cref{fig:sp_converging_31}). \cref{fig:4B_phasediagram} shows The dependence of the continuum $\mathcal{R}^{(4)}$ on $\eta$. We do not explicitly calculate two limiting cases $\eta\to0$ ($\Ep\to0$) and $\eta\to\infty$ ($\Es\to0$), but it is straightforward to locate them on the diagram. First as $\eta\to0$, the ground state reduces to a single $\uparrow\downarrow$-dimer and the two remaining $\uparrow$ spins are free spectators \cite{10.1063/1.1704798,Castella_1993}, hence $ E^{(\uparrow\uparrow\uparrow\downarrow)}/\Es\to 1$. On the other hand, as $\eta\to\infty$ ($\Es\to 0)$, the single $\downarrow$ spin decouples and the three $\uparrow$ spins form a $p$-wave cluster such that  $\Efour/\Ep=\Ethreep/\Ep\to 4$.

When $\eta=1$ the ratio is  $\mathcal{R}^{(4)}=10.002(3)$, which is equal to that of four identical fermions (\cref{sec:four+five_polarized}).  Notice that $4\uparrow$ contains six homogeneous $p$-wave pairs while $3\!\uparrow+1\!\downarrow$ contains an equal mix of $s$- and $p$-wave pairs. It is not obvious to us why matching the two-body binding energies yields the same ratio. Nevertheless, together with the previous calculations in the $2\!\uparrow+1\!\downarrow$ system, these results strongly indicate a correspondence between an $(N+1)$-body identical fermion system and a single fermion impurity coupled to $N$-body identical fermions when setting $\Es=\Ep$, i.e., only when $\eta =1$
\begin{equation}
    \frac{E^{(N\uparrow + 1\downarrow)}}{\Ep}=    \frac{E^{((N+1)\uparrow)}}{\Ep}.
\end{equation}

The four-body sector also admits the $2\!\uparrow+2\!\downarrow$ configuration.  Its lowest breakup channel is a pair of dimers, which is either two $s$-wave pairs or two $p$-wave pairs. The competition between two-dimer and genuine tetramer configurations has no three-body analogue. Extending this configuration to unequal masses or introducing two distinct $p$-wave couplings, $V_{2\uparrow}\neq V_{2\downarrow}$, are also of interest. We defer this analysis to future work.

%
%

\section{Conclusion  \label{sec:conclusion}}
In this paper, we revisit non-relativistic two-component fermion systems in one dimension interacting with an on-site interaction among the opposite spins and a nearest-neighbor interaction among the same spins. In the continuum, these interactions can be described as 
delta potential and delta-prime potential, respectively. The former is well-defined without regularization, but the latter would require one. 

A combined lattice and finite-length formulation offers a non-perturbative realization of this theory, enabling calculation of few-body spectra in a unified framework. The lattice spacing and the finite length together regulate the problem in the ultraviolet and the infrared. The couplings in each channel can be fixed non-perturbatively by a rank-one resolvent condition that reproduces the two-body binding energies exactly. It is worth emphasizing that the two channels are not equally accessible in a finite-length scheme. An arbitrary shallow $p$-wave dimer cannot be realized in a finite 1D box. Because the Hilbert space is finite-dimensional, the few-body problem can be solved exactly by exact diagonalization, with no approximation beyond the extrapolation in the limits $a\to0$ and $\Lphy\to\infty$. All results are expressed as dimensionless quantities, independent of the input mass, physical length of the box, or energy scale, and are therefore universal.

For two-component fermions interesting through the $s$-wave channel only, there is no genuine trimer state. The ground state consists of a dimer and a free particle, in agreement with the Gaudin–Yang solution. The slow convergence to the infinite-length limit observed in this case is a signature of the absence of binding, in contrast to the exponential convergence characteristic of a bound cluster.

For single-component fermions with the odd-wave interactions, the continuum extrapolations successfully reproduce the McGuire spectrum for up to five particles. This confirms the Bose-Fermi duality numerically. Furthermore, this spectrum is governed by the two-body couplings alone. Nevertheless, a three-body operator may still be introduced as a lattice improvement term, but it is not needed for renormalization.

Our main result is the universal functions $\mathcal{R}^{(3)}=\Ethrees/{\rm min}[\Es,\Ep]$ and $\mathcal{R}^{(4)}=\Efour/{\rm min}[\Es,\Ep]$ when both $s$- and $p$-wave interactions are present. We showed that both $\mathcal{R}^{(3)}$ and $\mathcal{R}^{(4)}$ only depend on the dimensionless ratio $\Ep/\Es$. The trimer is most deeply bound when  $\Ep=\Es$, attaining the McGuire exact value $\mathcal{R}^{(3)}=4$ and $\mathcal{R}^{(4)}=10$, respectively. Away from this point, the binding energy decreases monotonically.

The approach developed here extends naturally in several directions. For example, it will be interesting to change one or both couplings to be repulsive and study whether the bound state  survives. The second is to take into account the mass-imbalance effect since the $p$-wave coupling between two different species need not be the same once the particles are physically distinct. Moreover, mass imbalance introduces an additional parameter beyond $\eta$.  Another direction is the extension to finite density, in which in-medium two- and three-body energies could be obtained exactly rather than variationally, providing benchmarks for existing studies of the competition between pairing and tripling in this system. Finally, mapping the universal ratios onto realistic experimental parameters, for instance a ${}^{40}$K gas near coexisting $s$- and $p$-wave Feshbach resonances, would connect these results directly to measurable binding energies.

The Hilbert space of proposed systems grows significantly, so the required memory for these calculations scales up rapidly. For systems with relatively large sizes, more efficient implementations of the lattice methods are required, such as quantum Monte Carlo simulations. Quantum Monte Carlo simulations are extremely useful when perturbative techniques begin to fail, especially for quantum chromodynamics problems \cite{Singh_2019}.  However, it is difficult for such algorithms to deal with fermion systems due to the sign problem \cite {PhysRevLett.94.170201}. Various algorithms are proposed to solve the issue of 'sign problems', including the constrained path Monte Carlo \cite{PhysRevC.110.024002}, exact fermion Monte Carlo \cite{PhysRevC.110.024002}, and Worldline Monte Carlo simulations \cite{Singh_2019, PhysRevC.110.024002}. 

\section*{Acknowledgments}
{S.N. conceived and supervised the project.
Both Z.L. and S.N. carried out the numerical calculations and data analysis and drafted the manuscript.
S.N. guided the interpretation of the results and finalized the discussion. All authors contributed to manuscript revision. 

S.N. thanks R.~Springer and S.~Chandrasekharan for helpful discussion during the initial phase of the project. {We acknowledge the use of AI assistance, specifically Claude (Anthropic) \cite{claude-opus-5}, in developing code, refining the language and clarity
of this manuscript and providing citation to itself, before
our final rounds of manual review and revision by all
authors. The exact diagonalization code is verified for small lattice using \texttt{QuSpin}, a Python package for exact diagonalization \cite{SciPostPhys}.

Partial funding for this work was received from the Washington and Lee Lenfest Grant program and the Summer Research Scholars program.  This work used Jetstream2 at Indiana University through allocation PHY240288 from the Advanced Cyberinfrastructure Coordination Ecosystem: Services \& Support (ACCESS) program, which is supported by U.S. National Science Foundation grants \#2138259, \#2138286, \#2138307, \#2137603, and \#2138296.
}




\appendix

\section{Analytical expression of $I_{s}(E,a)$ in the continuum limit \label{app:Ilats}}
We start from the following equation
\begin{equation}
    I_{s}(E,a) = \frac{1}{L} \sum_{n=0}^{L-1} \frac{1}{E/2\epsilon(a) - 2[1 - \cos{(2\pi n / L)}]}.
\end{equation}
In the continuum limit, $a\to 0$ $L\to \infty$ while $La=\Lphy$ is fixed, we can use the Poisson summation formula to simplify the sum as
\begin{align}
 &\frac{1}{L} \sum_{n=0}^{L-1} \frac{1}{E/2\epsilon(a) - 2[1 - \cos{(2\pi n / L)}]}\nonumber\\
 =&\int_0^{2\pi} \frac{dq}{2\pi}\ f(q) +2\sum_{k=1}^{\infty}\int_0^{2\pi}\frac{dq}{2\pi }f(q)\cos(qkL),
\end{align}
where 
\begin{equation}
    f(q)=\frac{-1}{2+m|E|a^2-2\cos(q)}.
\end{equation}
We also assume that the two-body system has a binding energy of $-|E|$. Using the following integral identity,
\begin{equation}
    \int_0^{\pi}\frac{dx}{2\pi}\frac{\cos(nx)}{a-\cos(x)}=\frac{1}{2\sqrt{a^2-1}}\left(a-\sqrt{a^2-1}\right)^n 
\end{equation}
for $a^2>1$ and $n>0$, we can show that
all integrals are finite.
\begin{align}
 I_{s}(E,a)
 =\frac{-1}{2\sqrt{b^2-1}}\left[1+2\sum_{k=1}^\infty(b-\sqrt{b^2-1})^{kL}\right],
\end{align}
where $b\equiv 1+m|E|a^2/2 +\mathcal{O}(a^4)$. Neglecting terms of order $\mathcal{O}(a^4)$, it follows that
\begin{align}
 I_{s}(E,a)
 &=\frac{-1}{2a\sqrt{m|E|}}\left[1+2\sum_{k=1}^\infty(1-\sqrt{m|E|}a)^{\frac{k\Lphy}{a}}\right],\nonumber\\
 &=\frac{-1}{2a\sqrt{m|E|}}\left(1+2\sum_{k=1}^\infty e^{-k\sqrt{m|E|}\Lphy}\right).
\end{align}
We then use the identity $\displaystyle \sum_{k=1}^\infty e^{-kx}=\frac{e^{-x}}{1-e^{-x}}$ to rewrite the sum as 
\begin{align}
1+2 \sum_{k=1}^\infty e^{-k\sqrt{m|E|}\Lphy}&=\frac{1+e^{-\sqrt{m|E|}\Lphy}}{1-e^{-\sqrt{m|E|}\Lphy}}\nonumber\\
&= \coth\left(\frac{\sqrt{m|E|}\Lphy}{2}\right).
\end{align}

\section{Analytical expression of $I_p(E,a)$ in the continuum limit \label{app:Ilatp}}
We start from the following equation
\begin{equation}
    I_p(E,a) = \frac{1}{L} \sum_{n=0}^{L/2} \frac{2-2\cos(4\pi n/L)}{E/2\epsilon(a) - 2[1 - \cos{(2\pi n / L)}]}.
\end{equation}
In the continuum limit, $a\to 0$ $L\to \infty$ while $La=\Lphy$ is fixed, we can use the Poisson summation formula to simplify the sum as
\begin{align}
I_p(E,a)=\int_0^{\pi} \frac{dq}{2\pi}\ g(q) +2\sum_{k=1}^{\infty}\int_0^{\pi}\frac{dq}{2\pi }g(q)\cos(qkL),
\end{align}
where 
\begin{equation}
    g(q)=\frac{\cos(2q)-1}{1+m|E|a^2/2-\cos(q)}.
\end{equation}
We also assume that the two-body system has a binding energy of $-|E|$.  For simplicity, we will write $I_p\equiv I_1 +2\sum_{k=1}^\infty I_2^{(k)}$
\begin{align}
    I_1&= \frac{1}{2\sqrt{b^2-1}}\left[(b-\sqrt{b^2-1})^2 -1\right],\nonumber\\
    &=\frac{1}{2a\sqrt{m|E|}}(-2 a\sqrt{m|E|} + 2a^2 m|E|)\\
    &=-1+a\sqrt{m|E|}\nonumber
\end{align}
where $b\equiv 1+m|E|a^2/2 +\mathcal{O}(a^4)$, implying that $b^2-1=m|E|a^2+\mathcal{O}(a^4)$

\begin{align}
 I_2^{(k)}
 &=\frac{(b-\sqrt{b^2-1})^{kL}}{4\sqrt{b^2-1}}\left[\left(b-\sqrt{b^2-1} -\frac{1}{b-\sqrt{b^2-1}}\right)^{2}\right]\nonumber\\
 &=\frac{(1-\sqrt{m|E|}a)^{kL}}{4\sqrt{m|E|}a} 4m|E|a^2\nonumber\\
 &= \sqrt{m|E|}a(1-\sqrt{m|E|}a)^{kL}.
\end{align}

Neglecting terms of order $\mathcal{O}(a^2)$, it follows that

\begin{align}
    I_p=-1+a\sqrt{m|E|}\coth\left(\frac{\sqrt{m|E|}\Lphy}{2}\right)+\mathcal{O}(a^2)
\end{align}

\bibliography{ref}

@article{Tan_2008,
   title={Energetics of a strongly correlated Fermi gas},
   volume={323},
   ISSN={0003-4916},
   url={http://dx.doi.org/10.1016/j.aop.2008.03.004},
   DOI={10.1016/j.aop.2008.03.004},
   number={12},
   journal={Annals of Physics},
   publisher={Elsevier BV},
   author={Tan, Shina},
   year={2008},
   month=dec, pages={2952–2970} }

@article{Barth_2011,
   title={Tan relations in one dimension},
   volume={326},
   ISSN={0003-4916},
   url={http://dx.doi.org/10.1016/j.aop.2011.05.010},
   DOI={10.1016/j.aop.2011.05.010},
   number={10},
   journal={Annals of Physics},
   publisher={Elsevier BV},
   author={Barth, Marcus and Zwerger, Wilhelm},
   year={2011},
   month=oct, pages={2544–2565} }

@article{PhysRevA.103.043307,
  title = {Field-theoretical aspects of one-dimensional Bose and Fermi gases with contact interactions},
  author = {Sekino, Yuta and Nishida, Yusuke},
  journal = {Phys. Rev. A},
  volume = {103},
  issue = {4},
  pages = {043307},
  numpages = {16},
  year = {2021},
  month = {Apr},
  publisher = {American Physical Society},
  doi = {10.1103/PhysRevA.103.043307},
  url = {https://link.aps.org/doi/10.1103/PhysRevA.103.043307}
}

@article{Guo:2022cyw,
    author = "Guo, Yixin and Tajima, Hiroyuki",
    title = "{Stability against three-body clustering in one-dimensional spinless p-wave fermions}",
    eprint = "2208.03654",
    archivePrefix = "arXiv",
    primaryClass = "cond-mat.quant-gas",
    reportNumber = "RIKEN-iTHEMS-Report-22",
    doi = "10.1103/PhysRevA.106.043310",
    journal = "Phys. Rev. A",
    volume = "106",
    number = "4",
    pages = "043310",
    year = "2022"
}

@article{PhysRevA.97.061603,
  title = {Universal bound states of one-dimensional bosons with two- and three-body attractions},
  author = {Nishida, Yusuke},
  journal = {Phys. Rev. A},
  volume = {97},
  issue = {6},
  pages = {061603},
  numpages = {5},
  year = {2018},
  month = {Jun},
  publisher = {American Physical Society},
  doi = {10.1103/PhysRevA.97.061603},
  url = {https://link.aps.org/doi/10.1103/PhysRevA.97.061603}
}

@article{Klein:2018iqa,
    author = "Klein, Nico and Lee, Dean and Mei{\ss}ner, Ulf-G.",
    title = "{Lattice Improvement in Lattice Effective Field Theory}",
    eprint = "1807.04234",
    archivePrefix = "arXiv",
    primaryClass = "hep-lat",
    doi = "10.1140/epja/i2018-12676-1",
    journal = "Eur. Phys. J. A",
    volume = "54",
    number = "12",
    pages = "233",
    year = "2018"
}

@misc{claude-opus-5,
  author       = {{Anthropic}},
  title        = {Claude Opus 5},
  year         = {2026},
  howpublished = {\url{https://claude.ai}},
  note         = {Large language model. Accessed August 2026}
}

@article{Bloch_2008,
   title={Many-body physics with ultracold gases},
   volume={80},
   ISSN={1539-0756},
   url={http://dx.doi.org/10.1103/RevModPhys.80.885},
   DOI={10.1103/revmodphys.80.885},
   number={3},
   journal={Reviews of Modern Physics},
   publisher={American Physical Society (APS)},
   author={Bloch, Immanuel and Dalibard, Jean and Zwerger, Wilhelm},
   year={2008},
   month=July, pages={885–964} }

@article{Guo:2022xhc,
    author = "Guo, Yixin and Tajima, Hiroyuki",
    title = "{Competition between pairing and tripling in one-dimensional fermions with coexistent s- and p-wave interactions}",
    eprint = "2210.07042",
    archivePrefix = "arXiv",
    primaryClass = "cond-mat.quant-gas",
    reportNumber = "RIKEN-iTHEMS-Report-22",
    doi = "10.1103/PhysRevB.107.024511",
    journal = "Phys. Rev. B",
    volume = "107",
    number = "2",
    pages = "024511",
    year = "2023"
}

@article{Venu_2023,
   title={Unitary p-wave interactions between fermions in an optical lattice},
   volume={613},
   ISSN={1476-4687},
   url={http://dx.doi.org/10.1038/s41586-022-05405-6},
   DOI={10.1038/s41586-022-05405-6},
   number={7943},
   journal={Nature},
   publisher={Springer Science and Business Media LLC},
   author={Venu, Vijin and Xu, Peihang and Mamaev, Mikhail and Corapi, Frank and Bilitewski, Thomas and D’Incao, Jose P. and Fujiwara, Cora J. and Rey, Ana Maria and Thywissen, Joseph H.},
   year={2023},
   month=Jan, pages={262–267} }

@article{Symanzik:1983dc,
    author = "Symanzik, K.",
    title = "{Continuum Limit and Improved Action in Lattice Theories. 1. Principles and $\varphi^4$ Theory}",
    reportNumber = "DESY-83-016",
    doi = "10.1016/0550-3213(83)90468-6",
    journal = "Nucl. Phys. B",
    volume = "226",
    pages = "187--204",
    year = "1983"
}

@article{Symanzik:1983gh,
    author = "Symanzik, K.",
    title = "{Continuum Limit and Improved Action in Lattice Theories. 2. O(N) Nonlinear Sigma Model in Perturbation Theory}",
    reportNumber = "DESY-83-026",
    doi = "10.1016/0550-3213(83)90469-8",
    journal = "Nucl. Phys. B",
    volume = "226",
    pages = "205--227",
    year = "1983"
}

@article{PhysRevLett.125.263402,
  title = {Collisional Loss of One-Dimensional Fermions Near a $p$-Wave Feshbach Resonance},
  author = {Chang, Ya-Ting and Senaratne, Ruwan and Cavazos-Cavazos, Danyel and Hulet, Randall G.},
  journal = {Phys. Rev. Lett.},
  volume = {125},
  issue = {26},
  pages = {263402},
  numpages = {6},
  year = {2020},
  month = {Dec},
  publisher = {American Physical Society},
  doi = {10.1103/PhysRevLett.125.263402},
  url = {https://link.aps.org/doi/10.1103/PhysRevLett.125.263402}
}

@article{10.1063/1.1704798,
    author = {McGuire, J. B.},
    title = {Interacting Fermions in One Dimension. II. Attractive Potential},
    journal = {Journal of Mathematical Physics},
    volume = {7},
    number = {1},
    pages = {123-132},
    year = {1966},
    month = {01},
    issn = {0022-2488},
    doi = {10.1063/1.1704798},
    url = {https://doi.org/10.1063/1.1704798}
}

@article{Castella_1993,
   title={Exact calculation of spectral properties of a particle interacting with a one-dimensional fermionic system},
   volume={47},
   ISSN={1095-3795},
   url={http://dx.doi.org/10.1103/PhysRevB.47.16186},
   DOI={10.1103/physrevb.47.16186},
   number={24},
   journal={Physical Review B},
   publisher={American Physical Society (APS)},
   author={Castella, H. and Zotos, X.},
   year={1993},
   month=June, pages={16186–16193} }

@article{RevModPhys.82.1225,
  title = {Feshbach resonances in ultracold gases},
  author = {Chin, Cheng and Grimm, Rudolf and Julienne, Paul and Tiesinga, Eite},
  journal = {Rev. Mod. Phys.},
  volume = {82},
  issue = {2},
  pages = {1225--1286},
  numpages = {0},
  year = {2010},
  month = {Apr},
  publisher = {American Physical Society},
  doi = {10.1103/RevModPhys.82.1225},
  url = {https://link.aps.org/doi/10.1103/RevModPhys.82.1225}
}

@article{jackson2023emergent,
  title={Emergent s-wave interactions between identical fermions in quasi-one-dimensional geometries},
  author={Jackson, Kenneth G and Dale, Colin J and Maki, Jeff and Xie, Kevin GS and Olsen, Ben A and Ahmed-Braun, Denise JM and Zhang, Shizhong and Thywissen, Joseph H},
  journal={Physical Review X},
  volume={13},
  number={2},
  pages={021013},
  year={2023},
  publisher={APS}
}

@article{Guo:2023cfq,
    author = "Guo, Yixin and Tajima, Hiroyuki",
    title = "{Cooper pairing and tripling in one-dimensional spinless fermions with attractive two- and three-body forces}",
    eprint = "2308.04737",
    archivePrefix = "arXiv",
    primaryClass = "cond-mat.quant-gas",
    reportNumber = "RIKEN-iTHEMS-Report-23",
    doi = "10.1103/PhysRevA.108.043303",
    journal = "Phys. Rev. A",
    volume = "108",
    number = "4",
    pages = "043303",
    year = "2023"
}

@article{Cazalilla_2011,
   title={One dimensional bosons: From condensed matter systems to ultracold gases},
   volume={83},
   ISSN={1539-0756},
   url={http://dx.doi.org/10.1103/RevModPhys.83.1405},
   DOI={10.1103/revmodphys.83.1405},
   number={4},
   journal={Reviews of Modern Physics},
   publisher={American Physical Society (APS)},
   author={Cazalilla, M. A. and Citro, R. and Giamarchi, T. and Orignac, E. and Rigol, M.},
   year={2011},
   month=Dec, pages={1405–1466} }

@article{PhysRevA.97.061605,
  title = {One-dimensional three-boson problem with two- and three-body interactions},
  author = {Guijarro, G. and Pricoupenko, A. and Astrakharchik, G. E. and Boronat, J. and Petrov, D. S.},
  journal = {Phys. Rev. A},
  volume = {97},
  issue = {6},
  pages = {061605},
  numpages = {5},
  year = {2018},
  month = {Jun},
  publisher = {American Physical Society},
  doi = {10.1103/PhysRevA.97.061605},
  url = {https://link.aps.org/doi/10.1103/PhysRevA.97.061605}
}

@article{PhysRevC.110.024002,
  title = {Worldline Monte Carlo method for few-body nuclear physics},
  author = {Chandrasekharan, Shailesh and Nguyen, Son T. and Richardson, Thomas R.},
  journal = {Phys. Rev. C},
  volume = {110},
  issue = {2},
  pages = {024002},
  numpages = {21},
  year = {2024},
  month = {Aug},
  publisher = {American Physical Society},
  doi = {10.1103/PhysRevC.110.024002},
  url = {https://link.aps.org/doi/10.1103/PhysRevC.110.024002}
}

@article{Singh_2019,
   title={Few-body physics on a spacetime lattice in the worldline approach},
   volume={99},
   ISSN={2470-0029},
   url={http://dx.doi.org/10.1103/PhysRevD.99.074511},
   DOI={10.1103/physrevd.99.074511},
   number={7},
   journal={Physical Review D},
   publisher={American Physical Society (APS)},
   author={Singh, Hersh and Chandrasekharan, Shailesh},
   year={2019},
   month=apr }

@article{Zhu:2019dho,
    author = "Zhu, Shangguo and Tan, Shina",
    title = {{$d$-dimensional L{\"u}scher's formula and the near-threshold three-body states in a finite volume}},
    eprint = "1905.05117",
    archivePrefix = "arXiv",
    primaryClass = "nucl-th",
    month = "5",
    year = "2019",
    journal = {arXiv e-prints}
}

@Article{SciPostPhys,
	title={{QuSpin: a Python package for dynamics and exact diagonalisation of quantum many body systems. Part II: bosons, fermions and higher spins}},
	author={Phillip Weinberg and Marin Bukov},
	journal={SciPost Phys.},
	volume={7},
	pages={020},
	year={2019},
	publisher={SciPost},
	doi={10.21468/SciPostPhys.7.2.020},
	url={https://scipost.org/10.21468/SciPostPhys.7.2.020},
}

@article{Korber:2019cuq,
    author = {K{\"o}rber, Christopher and Berkowitz, Evan and Luu, Thomas},
    title = "{Renormalization of a Contact Interaction on a Lattice}",
    eprint = "1912.04425",
    archivePrefix = "arXiv",
    primaryClass = "hep-lat",
    month = "12",
    year = "2019",
    journal = {arXiv e-prints}
}

@article{Yang:1967bm,
    author = "Yang, Chen-Ning",
    title = "{Some exact results for the many body problems in one dimension with repulsive delta function interaction}",
    doi = "10.1103/PhysRevLett.19.1312",
    journal = "Phys. Rev. Lett.",
    volume = "19",
    pages = "1312--1314",
    year = "1967"
}

@article{GAUDIN196755,
title = {Un systeme a une dimension de fermions en interaction},
journal = {Physics Letters A},
volume = {24},
number = {1},
pages = {55-56},
year = {1967},
issn = {0375-9601},
doi = {https://doi.org/10.1016/0375-9601(67)90193-4},
url = {https://www.sciencedirect.com/science/article/pii/0375960167901934},
author = {M. Gaudin}
}

@article{PhysRevLett.94.170201,
  title = {Computational Complexity and Fundamental Limitations to Fermionic Quantum Monte Carlo Simulations},
  author = {Troyer, Matthias and Wiese, Uwe-Jens},
  journal = {Phys. Rev. Lett.},
  volume = {94},
  issue = {17},
  pages = {170201},
  numpages = {4},
  year = {2005},
  month = {May},
  publisher = {American Physical Society},
  doi = {10.1103/PhysRevLett.94.170201},
  url = {https://link.aps.org/doi/10.1103/PhysRevLett.94.170201}
}

@book{Montvay_Münster_1994, 
place={Cambridge}, 
series={Cambridge Monographs on Mathematical Physics}, 
title={Quantum Fields on a Lattice}, 
publisher={Cambridge University Press}, 
author={Montvay, Istvan and Münster, Gernot}, 
year={1994}, 
collection={Cambridge Monographs on Mathematical Physics}
}

@article{Lenci2010,
author={Lenci, Marco},
title={On Infinite-Volume Mixing},
journal={Communications in Mathematical Physics},
year={2010},
month={Sep},
day={01},
volume={298},
number={2},
pages={485-514},
issn={1432-0916},
doi={10.1007/s00220-010-1043-6},
url={https://doi.org/10.1007/s00220-010-1043-6}
}

@article{PhysRev.135.A1172,
  title = {Finite Superconductors and their Infinite Volume Limit},
  author = {Henley, E. M. and Kennedy, R. C. and Wilets, L.},
  journal = {Phys. Rev.},
  volume = {135},
  issue = {5A},
  pages = {A1172--A1174},
  numpages = {0},
  year = {1964},
  month = {Aug},
  publisher = {American Physical Society},
  doi = {10.1103/PhysRev.135.A1172},
  url = {https://link.aps.org/doi/10.1103/PhysRev.135.A1172}
}

@book{Griffiths_Schroeter_2018, 
place={Cambridge}, 
edition={3}, 
title={Introduction to Quantum Mechanics}, 
publisher={Cambridge University Press}, 
author={Griffiths, David J. and Schroeter, Darrell F.}, 
year={2018}
}

@article{Liao_2010,
   title={Spin-imbalance in a one-dimensional Fermi gas},
   volume={467},
   ISSN={1476-4687},
   url={http://dx.doi.org/10.1038/nature09393},
   DOI={10.1038/nature09393},
   number={7315},
   journal={Nature},
   publisher={Springer Science and Business Media LLC},
   author={Liao, Yean-an and Rittner, Ann Sophie C. and Paprotta, Tobias and Li, Wenhui and Partridge, Guthrie B. and Hulet, Randall G. and Baur, Stefan K. and Mueller, Erich J.},
   year={2010},
   month=sep, pages={567–569} }

@article{Pagano_2014,
   title={A one-dimensional liquid of fermions with tunable spin},
   volume={10},
   ISSN={1745-2481},
   url={http://dx.doi.org/10.1038/nphys2878},
   DOI={10.1038/nphys2878},
   number={3},
   journal={Nature Physics},
   publisher={Springer Science and Business Media LLC},
   author={Pagano, Guido and Mancini, Marco and Cappellini, Giacomo and Lombardi, Pietro and Schäfer, Florian and Hu, Hui and Liu, Xia-Ji and Catani, Jacopo and Sias, Carlo and Inguscio, Massimo and Fallani, Leonardo},
   year={2014},
   month=feb, pages={198–201} }

@article{Wenz_2013,
   title={From Few to Many: Observing the Formation of a Fermi Sea One Atom at a Time},
   volume={342},
   ISSN={1095-9203},
   url={http://dx.doi.org/10.1126/science.1240516},
   DOI={10.1126/science.1240516},
   number={6157},
   journal={Science},
   publisher={American Association for the Advancement of Science (AAAS)},
   author={Wenz, A. N. and Zürn, G. and Murmann, S. and Brouzos, I. and Lompe, T. and Jochim, S.},
   year={2013},
   month=oct, pages={457–460} }

@article{He_2016,
   title={Universal properties of Fermi gases in one dimension},
   volume={94},
   ISSN={2469-9934},
   url={http://dx.doi.org/10.1103/PhysRevA.94.031604},
   DOI={10.1103/physreva.94.031604},
   number={3},
   journal={Physical Review A},
   publisher={American Physical Society (APS)},
   author={He, Wen-Bin and Chen, Yang-Yang and Zhang, Shizhong and Guan, Xi-Wen},
   year={2016},
   month=sep }

@article{Pecak:2017mqz,
    author = "P{\k{e}}cak, Daniel and Dehkharghani, Amin S. and Zinner, Nikolaj T. and Sowi{\'n}ski, Tomasz",
    title = "{Four fermions in a one-dimensional harmonic trap: Accuracy of a variational-ansatz approach}",
    eprint = "1703.08720",
    archivePrefix = "arXiv",
    primaryClass = "cond-mat.quant-gas",
    doi = "10.1103/PhysRevA.95.053632",
    journal = "Phys. Rev. A",
    volume = "95",
    pages = "053632",
    year = "2017"
}

@article{Tanaka_2022,
   title={Bulk Viscosity of Dual Bose and Fermi Gases in One Dimension},
   volume={129},
   ISSN={1079-7114},
   url={http://dx.doi.org/10.1103/PhysRevLett.129.200402},
   DOI={10.1103/physrevlett.129.200402},
   number={20},
   journal={Physical Review Letters},
   publisher={American Physical Society (APS)},
   author={Tanaka, Tomohiro and Nishida, Yusuke},
   year={2022},
   month=Nov }

@article{Guan_2013,
   title={Fermi gases in one dimension: From Bethe ansatz to experiments},
   volume={85},
   ISSN={1539-0756},
   url={http://dx.doi.org/10.1103/RevModPhys.85.1633},
   DOI={10.1103/revmodphys.85.1633},
   number={4},
   journal={Reviews of Modern Physics},
   publisher={American Physical Society (APS)},
   author={Guan, Xi-Wen and Batchelor, Murray T. and Lee, Chaohong},
   year={2013},
   month=nov, pages={1633–1691} }

@article{Luscher:1985dn,
    author = "Luscher, M.",
    title = "{Volume Dependence of the Energy Spectrum in Massive Quantum Field Theories. 1. Stable Particle States}",
    reportNumber = "DESY-85-144",
    doi = "10.1007/BF01211589",
    journal = "Commun. Math. Phys.",
    volume = "104",
    pages = "177",
    year = "1986"
}

@article{Luscher:1986pf,
    author = "Luscher, M.",
    title = "{Volume Dependence of the Energy Spectrum in Massive Quantum Field Theories. 2. Scattering States}",
    reportNumber = "DESY-86-034",
    doi = "10.1007/BF01211097",
    journal = "Commun. Math. Phys.",
    volume = "105",
    pages = "153--188",
    year = "1986"
}

@article{Cheon_1998,
   title={Realizing discontinuous wave functions with renormalized short-range potentials},
   volume={243},
   ISSN={0375-9601},
   url={http://dx.doi.org/10.1016/S0375-9601(98)00188-1},
   DOI={10.1016/s0375-9601(98)00188-1},
   number={3},
   journal={Physics Letters A},
   publisher={Elsevier BV},
   author={Cheon, Taksu and Shigehara, T},
   year={1998},
   month=jun, pages={111–116} }

@book{baxter2007exactly,
  title={Exactly Solved Models in Statistical Mechanics},
  author={Baxter, R.J.},
  isbn={9780486462714},
  lccn={2007037510},
  series={Dover books on physics},
  url={https://books.google.com/books?id=G3owDULfBuEC},
  year={2007},
  publisher={Dover Publications}
}

@book{giamarchi2004quantum,
  title={Quantum Physics in One Dimension},
  author={Giamarchi, T.},
  isbn={9780198525004},
  lccn={2004299020},
  series={International Series of Monographs on Physics},
  url={https://books.google.com/books?id=1MwTDAAAQBAJ},
  year={2004},
  publisher={Clarendon Press}
}

@article{PhysRevLett.94.210401,
  title = {Confinement Induced Molecules in a 1D Fermi Gas},
  author = {Moritz, Henning and St\"oferle, Thilo and G\"unter, Kenneth and K\"ohl, Michael and Esslinger, Tilman},
  journal = {Phys. Rev. Lett.},
  volume = {94},
  issue = {21},
  pages = {210401},
  numpages = {4},
  year = {2005},
  month = {Jun},
  publisher = {American Physical Society},
  doi = {10.1103/PhysRevLett.94.210401},
  url = {https://link.aps.org/doi/10.1103/PhysRevLett.94.210401}
}

@article{O_Hara_2002,
   title={Observation of a Strongly Interacting Degenerate Fermi Gas of Atoms},
   volume={298},
   ISSN={1095-9203},
   url={http://dx.doi.org/10.1126/science.1079107},
   DOI={10.1126/science.1079107},
   number={5601},
   journal={Science},
   publisher={American Association for the Advancement of Science (AAAS)},
   author={O’Hara, K. M. and Hemmer, S. L. and Gehm, M. E. and Granade, S. R. and Thomas, J. E.},
   year={2002},
   month=dec, pages={2179–2182} }

@article{PhysRevLett.102.230402,
  title = {Observation of Fermi Polarons in a Tunable Fermi Liquid of Ultracold Atoms},
  author = {Schirotzek, Andr\'e and Wu, Cheng-Hsun and Sommer, Ariel and Zwierlein, Martin W.},
  journal = {Phys. Rev. Lett.},
  volume = {102},
  issue = {23},
  pages = {230402},
  numpages = {4},
  year = {2009},
  month = {Jun},
  publisher = {American Physical Society},
  doi = {10.1103/PhysRevLett.102.230402},
  url = {https://link.aps.org/doi/10.1103/PhysRevLett.102.230402}
}

@article{McGuire_1964,
    author = "McGuire, J. B.",
    title = "{Study of Exactly Soluble One-Dimensional N-Body Problems}",
    doi = "10.1063/1.1704156",
    journal = "J. Math. Phys.",
    volume = "5",
    number = "5",
    pages = "622--636",
    year = "1964"
}

@article{Cheon:1998iy,
    author = "Cheon, Taksu and Shigehara, T.",
    title = "{Fermion boson duality of one-dimensional quantum particles with generalized contact interaction}",
    eprint = "quant-ph/9806041",
    archivePrefix = "arXiv",
    doi = "10.1103/PhysRevLett.82.2536",
    journal = "Phys. Rev. Lett.",
    volume = "82",
    pages = "2536--2539",
    year = "1999"
}

@ARTICLE{1963PhRv..130.1605L,
       author = {{Lieb}, Elliott H. and {Liniger}, Werner},
        title = "{Exact Analysis of an Interacting Bose Gas. I. The General Solution and the Ground State}",
      journal = {Physical Review},
         year = 1963,
        month = may,
       volume = {130},
       number = {4},
        pages = {1605-1616},
          doi = {10.1103/PhysRev.130.1605},
       adsurl = {https://ui.adsabs.harvard.edu/abs/1963PhRv..130.1605L}
}

@article{takahashi1971one,
  title={One-dimensional electron gas with delta-function interaction at finite temperature},
  author={Takahashi, Minoru},
  journal={Progress of Theoretical Physics},
  volume={46},
  number={5},
  pages={1388--1406},
  year={1971},
  publisher={Oxford University Press}
}
\end{document}